\documentclass[fleqn,usenatbib]{mnrasmod}
\usepackage[T1]{fontenc}
\usepackage{ae,aecompl}
\usepackage{graphicx}	% Including figure files
\usepackage{amsmath}	% Advanced maths commands
\usepackage{amssymb}	% Extra maths symbols
\usepackage{dsfont}
\usepackage{newtxtext,newtxmath}
\usepackage{natbib}

\title[Multifractal Analysis of Hi-GAL maps]{Multifractal Analysis of the Interstellar Medium. First
application to Hi-GAL Observations}

\author[D. Elia et al.]{Davide Elia,$^{1}$\thanks{E-mail: davide.elia@iaps.inaf.it}
Francesco Strafella,$^{2,3}$
Sami Dib,$^{4,5}$
Nicola Schneider,$^{6}$
Patrick Hennebelle,$^{7,8,9}$
\newauthor
Stefano Pezzuto,$^{1}$
Sergio Molinari,$^{1}$
Eugenio Schisano,$^{1}$
Sarah E. Jaffa$^{10}$
\\
$^{1}$Istituto di Astrofisica e Planetologia Spaziali, INAF, Via Fosso del Cavaliere 100, Roma, 00133, Italy\\
$^{2}$Dipartimento di Matematica e Fisica `Ennio De Giorgi', Universit\`{a} del Salento, CP 193, I-73100 Lecce, Italy\\
$^{3}$INFN Sezione di Lecce, CP 193, I-73100 Lecce, Italy\\
$^{4}$Niels Bohr International Academy, Niels Bohr Institute, Blegdamsvej 17, DK-2100, Copenhagen, Denmark\\
$^{5}$Max-Planck Institute f\"{u}r Astronomie, K\"{o}nigstuhl 17, 69117, Heidelberg, Germany\\
$^{6}$I. Physik. Institut, University of Cologne, D-50937 Cologne, Germany\\
$^{7}$IRFU, CEA, Universit\'e Paris-Saclay, 91191 Gif-sur-Yvette, France \\
$^{8}$Universit\'e Paris Diderot, AIM, Sorbonne Paris Cit\'e, CEA, CNRS, 91191 Gif-sur-Yvette, France \\
$^{9}$LERMA (UMR CNRS 8112), Ecole Normale Sup\'erieure, 75231 Paris Cedex, France\\
$^{10}$School of Physics and Astronomy, Cardiff University, Cardiff CF24 3AA, Wales, UK 0000-0002-6711-6345\\ }

\date{Accepted XXX. Received YYY; in original form ZZZ}

\pubyear{2017}

\begin{document}
\label{firstpage}
\pagerange{\pageref{firstpage}--\pageref{lastpage}}
\maketitle

\begin{abstract}
The multifractal geometry remains an under-exploited approach to describe and 
quantify the large-scale structure of interstellar clouds. In this paper, the typical tools 
of multifractal analysis are applied to \textit{Herschel} far-infrared (70-500~$\mu$m) dust
continuum maps, which represent an ideal case of study. This dust component is a relatively 
optically thin tracer at these wavelengths and the size in pixel of the maps is well 
suitable for this statistical analysis.
We investigate the so-called multifractal spectrum and generalised fractal dimension of
six Hi-GAL regions in the third Galactic quadrant. We measure systematic variations 
of the spectrum at increasing wavelength, which generally correspond to an increasing complexity 
of the image, and we observe peculiar behaviours of the investigated fields, strictly
related to the presence of high-emission regions, which in turn are connected to star 
formation activity. The same analysis is applied to synthetic column density maps, generated  
from numerical turbulent molecular cloud models and from fractal Brownian motion (fBm), allowing 
for the confrontation of the observations with models with well controlled physical parameters.
This comparison shows that, in terms of multifractal descriptors of the 
structure, fBm images exhibit a very different, quite unrealistic behaviour when compared with Hi-GAL
observations, whereas the numerical simulations appear in some cases (depending on the specific model) 
more similar to the observations. Finally, the link between mono-fractal 
parameters (commonly found in the literature) and multifractal indicators is investigated: the former
appear to be only partially connected with the latter, with the multifractal analysis
offering a more extended and complete characterization of the cloud structure.
\end{abstract}


%%%%%%%%%%%%%%%%%%%%%%%%%%%%%%%%%%%%%%%%%%%%%%%%%%

%%%%%%%%%%%%%%%%% BODY OF PAPER %%%%%%%%%%%%%%%%%%

\newcommand{\imasizeblue}{1690}
\newcommand{\angsizeblue}{5408.00}
\newcommand{\divisorsblue}{2, 5, 10, 13, 26, 65, 130, 169, 338, 845}
\newcommand{\ndivisorsblue}{10}
\newcommand{\imasizered}{1200}
\newcommand{\angsizered}{5400.00}
\newcommand{\divisorsred}{2, 3, 4, 5, 6, 8, 10, 12, 15, 16, 20, 24, 25, 30, 40, 48, 50, 60, 75, 80, 100, 120, 150, 200, 240, 300, 400, 600}
\newcommand{\ndivisorsred}{28}
\newcommand{\imasizePSW}{900}
\newcommand{\angsizePSW}{5400.00}
\newcommand{\divisorsPSW}{2, 3, 4, 5, 6, 9, 10, 12, 15, 18, 20, 25, 30, 36, 45, 50, 60, 75, 90, 100, 150, 180, 225, 300, 450}
\newcommand{\ndivisorsPSW}{25}
\newcommand{\imasizePMW}{675}
\newcommand{\angsizePMW}{5400.00}
\newcommand{\divisorsPMW}{3, 5, 9, 15, 25, 27, 45, 75, 135, 225}
\newcommand{\ndivisorsPMW}{10}
\newcommand{\imasizePLW}{470}
\newcommand{\angsizePLW}{5405.00}
\newcommand{\divisorsPLW}{2, 5, 10, 47, 94, 235}
\newcommand{\ndivisorsPLW}{6}   %%% generated by imagestats.pro

\section{Introduction}\label{intro}
\defcitealias{eli14}{Paper~I}
The complex morphology of Galactic interstellar clouds generally eludes a description simply based on the 
typical shapes of Euclidean geometry \citep[e.g.,][]{sca93,elm96,pfe96}. The mostly self-similar 
appearance of the interstellar medium (ISM), especially at scales $L \gtrsim 1$~pc is generally 
thought to be the result of turbulence \citep{elm95,elm04,dib05,kru14}, 
because of the intrinsic self-similarity of this phenomenon, triggered by the very high value of the
Reynolds number in these environments. The fractal geometry is largely invoked to provide a quantitative
characterization of the ISM morphology, as it can be deduced from far-infrared (FIR) or sub-millimetre maps,
through fractal descriptors \citep[][hereafter \citetalias{eli14}]{stu98,ben01,sch11,sun18,eli14} and 
comparing these quantities with theoretical expectations has 
two important advantages. On the one hand, a comparison between models and observations gives 
indications about the kind of turbulence that is prevalent in the observed cloud and, 
on the other hand, the obtained fractal parameters can be used as further constraints for future 
simulations. 
In this respect, fractal or, in a broad sense, statistical analysis techniques are applied also to 
numerical simulations of turbulent clouds, to make possible a comparison between model and ISM morphological
properties. Three-dimensional magneto-hydrodynamic simulations have been characterised in terms of probability 
density function, structure function and power spectrum \citep{kow07,kri07}, fractal dimension \citep{fed09}, 
or dendrograms \citep{burk13}.

It should also be emphasised that natural fractal structures, such as ISM, are self-similar only 
in a statistical sense (\textit{stochastic fractals}) and show complexity only over a finite 
range of scales. Identifying such a range, which should correspond to the scales actually involved 
by the turbulent energy cascade, can give important information on the injection and dissipation 
scale of turbulence \citepalias[see][and references therein]{eli14}. Interestingly,
the latter is expected to correspond to the characteristic size of gravity-dominated
structures (filaments and cores) relevant for star formation, as highlighted, e.g.,
by \citet{fal04} and \citet{sch11} \citep[for more recent analysis of the turbulence-regulated 
star formation, see also][]{bur17,moc17}.

This (mono-)fractal approach, however, still underlies a certain degree of degeneracy: as said 
above, a large number of natural objects exhibit properties of self-similarity, but can not 
be described as perfect fractals, especially because such a description is based on a sole 
parameter, namely the fractal dimension. In contrast, \textit{Multifractal geometry} provides a 
more suited mathematical framework for detecting and identifying complex local structure, and 
for describing local singularities. It is invoked in the fields of economy, medicine, hydrography, 
environmental sciences, and physics (e.g. in studying turbulent flows). In astrophysics, it has 
been used, for example, in the study of the distribution of galaxies \citep{man89,bor93,pan00,del06}, 
gamma-ray burst time series \citep{mer95}, and solar activity \citep{wu15,cad16,mar17}.
 
The first application of the multifractal approach to characterize the ISM structure was 
carried out by \citet{cha01}, who analyzed 13 \emph{IRAS} maps (at 60 and 100~$\mu$m) 
taken from Chamaeleon-Musca, R~Corona Australis and Scorpius-Ophiucus star forming regions, 
all located at distance $d<160$~pc from the Sun, but spanning a relatively wide range 
of different conditions of star formation. A multifractal behaviour of the investigated 
clouds was revealed over the range 0.4-4~pc, and a possible relation with underlying 
turbulent cascades and hierarchical structure was found. While the multifractal properties 
of a cloud can be directly related to its geometry, a direct relation with the properties 
of the star formation is not found by these authors; this link has been established indirectly
for the first time by \citet{tas07} using global parameters of external galaxies. 
Finally, \citet{kha06} studied the multifractal structure of Galactic atomic hydrogen through 
wavelet transform techniques, discussing arm vs inter-arm differences.

As already pointed out in \citetalias{eli14}, the advent of \textit{Herschel} \citep{pil10} 
offered an extraordinary chance for studying the morphology of the cold Galactic ISM, thanks 
to the combination of several favourable conditions and features: $i$) the spectral coverage 
of \textit{Herschel} photometric surveys ($70-500~\mu$m) encompasses the peak of the continuum 
emission of cold dust ($T \leq 50$~K), with the dust getting optically thinner at increasing 
wavelengths, which helps revealing the structure of dense clouds with unprecedented accuracy; 
$ii$) the angular resolution of \textit{Herschel} photometric observations ($6\arcsec-36\arcsec$)
is better or comparable with that of the most recent CO surveys of the ISM 
\citep[e.g.,][]{jac06,bur13,sch17}, and their dynamic range is so large 
\citep[more than two orders of magnitude, e.g.,][]{mol16} that the obtained 2-D picture of the 
ISM structure generally turns out to be highly 
detailed and reliable; $iii$) large \textit{Herschel} photometric surveys produced a huge amount 
of data, corresponding to a wide variety of Galactic locations, physical conditions, and star 
formation modes to be compared; $iv$) thanks to good angular resolution and excellent mapping 
capabilities, single \textit{Herschel} maps represent highly suitable data sets for pixel 
statistical analysis, namely techniques aimed at deriving fractal properties of maps starting from 
intensity values in single pixels. This improvement can be appreciated, for instance, by comparing 
how much the size in pixels of the analysed maps increased from the pioneering work of, e.g., 
\citet{stu98} (several tens or a few hundred pixels) to that of \citetalias{eli14} (up to a 
few thousands of pixels).

Despite this potential, the promising approach of \citet{cha01} for describing the structure of the 
ISM through multifractal parameters has not been yet applied to \textit{Herschel} maps. With this 
paper we intend to fill this gap. Furthermore, choosing the same fields already analysed
through mono-fractal descriptors in \citetalias{eli14}, we explore possible relations between
mono- and multifractal parameters for any investigated field. Finally, analysing with the same 
techniques also column density maps obtained from numerical simulations of insterstellar turbulence, 
we search for possible recipes
connecting the quantitative description provided by multifractal tools and the underlying physics
of the analysed regions. 

The paper is organized as follows: in Section~\ref{fields} the analysed maps, both observational 
and synthetic, are presented. In Section~\ref{multi} basics of multifractal geometry 
are introduced, together with a description of the tools used in the rest of the paper. In 
Section~\ref{results} the results of the multifractal analysis of the aforementioned data sets
are reported and discussed by means of specific diagnostics, while our conclusions are given in 
Section~\ref{conclusions}. Additional details and discussion
of the method are reported in the Appendices~\ref{fbmappend}, \ref{simevol}, and \ref{dvarapp}.

\section{Analysed fields}\label{fields}

%The targets of our analysis are six Hi-GAL fields of the third Galactic quadrant, four
%of which have been already analysed in \citetalias{eli14}, and a set of fractional Brownian motion
%images generated as reference. 

\subsection{Hi-GAL maps}\label{higal} 
The observational sets chosen for this analysis are six $1.5^\circ \times 1.5^\circ$
fields extracted from the Hi-GAL programme 
\citep{mol10}, the photometric survey of the entire Galactic plane carried out at the 70, 160, 250, 350, 
and 500~$\mu$m wavelengths with the two cameras PACS \citep{pog10} and SPIRE \citep{gri10}. The 
original $\sim 2.2^\circ \times 2.2^\circ$ Hi-GAL tiles obtained in the five aforementioned bands have 
a nominal resolution\footnote{Actually, due to the PACS data co-adding on-board Herschel, the resulting 
point-spread functions are elongated along the scan direction, with a measured size of 5.8\arcsec 
$\times$ 12.1\arcsec at 70~$\mu$m, and 11.4\arcsec $\times$ 13.4\arcsec at 160~$\mu$m, respectively 
\citep{lut12}.} of 5.6\arcsec, 11.3\arcsec, 17.6\arcsec, 23.9\arcsec, and 35.1\arcsec, and a pixel size of 
3.2\arcsec, 4.5\arcsec, 6\arcsec, 8\arcsec, and 11.5\arcsec, respectively \citep{mol16}.

The fields were chosen in the third Galactic quadrant \citep{eli13}: the central four, designated as 
$\ell217$, $\ell220$, $\ell222$, $\ell224$ (from the original names of Hi-GAL the tiles they were 
extracted from), respectively, are already described (including data reduction and map making details) 
in \citet{eli13} and in \citetalias{eli14}. These fields were selected based on the following considerations:
\begin{enumerate}
\item Ideally, each selected field should contain emission from a single Galactic component, i.e. from 
dust distributed within a limited range of heliocentric distances. In this portion of the Galactic 
plane, NANTEN data \citep{oni05,eli13} show a simple structure of the velocity field 
\citepalias[][their Figure~1]{eli14}, so that each of the chosen Hi-GAL fields can be associated with 
a single predominant velocity component, i.e. a single coherent cloud.
\item The clipping of original tiles must be performed on the area observed by both PACS and SPIRE, 
choosing the same box for all the five bands, in order to analyse the same area of the sky at 
all wavelengths, in the limit of pixelation.   
\item The clipped maps are chosen to be square-shaped and must be as large as possible, in order to 
increase the statistical relevance of the results and expand the range of spatial scales that are probed. 
\end{enumerate} 
In addition to these requirements, for this work we also impose that 
\begin{enumerate}
 \setcounter{enumi}{3}
\item The size of the analysed maps must be a number of pixels characterised by a large enough number 
of integer factors, for the reasons that will be explained in Section~\ref{mfscalc}.
\end{enumerate} 
About cropping the maps, it is difficult to find a common angular size, for all the five bands, which 
contextually corresponds to sizes (in pixels) satisfying the aforementioned property. For this reason we  
slightly adjusted the number of pixels composing the maps at each band, finding a final set of frame sizes in 
pixels approximately corresponding to $1.5^{\circ}$ (depending on the wavelength, see Table~\ref{imatab}). 
As a consequence, the maps analysed in this work turn out to be slightly smaller than those for 
which the fractal dimension was derived in \citetalias{eli14}, making it necessary to re-compute it (see 
Appendix~\ref{dvarapp}) in view of a comparison with multi-fractal properties (Section~\ref{dqanalysis}). 

\begin{table}
\caption{Geometric properties of analysed frames.}
\label{imatab}
\begin{tabular}{lccc}
\hline
Map type & Frame size & $N_\mathrm{divisors}$ & Frame angular size \\
         & pixel &   & $\arcsec$ \\
\hline
70~$\mu$m &   1690 &  10 &    5408 \\
160~$\mu$m &   1200 &  28 &    5400 \\
250~$\mu$m &    900 &  25 &    5400 \\
350~$\mu$m &    675 &  10 &    5400 \\
500~$\mu$m &    470 &   6 &    5405 \\
fBm & 1000 &           14 & ...\\
Mod.: Solenoidal & 256 &            7 & ...\\
Mod.: Compressive & 256 &            7 & ...\\
Mod.: Hydrodynamical & 2048 &           10 & ...\\
Mod.: High-magnetization & 2048 &           10 & ...\\
\hline
\end{tabular}
\end{table}

In Table~\ref{imatab} we report for each band the size of the maps in pixels, and the 
corresponding number of factors\footnote{For example, at 350~$\mu$m the chosen map size is 
$\imasizePMW\times \imasizePMW$~pixels. Except for 1 and itself, the side 
\imasizePMW~has the following integer divisors:~\divisorsPMW, which therefore allow to perfectly 
cover the map with \ndivisorsPMW~possible box grids, and so on.} and angular scale in arcseconds. 
\begin{figure*}
\includegraphics[width=17cm]{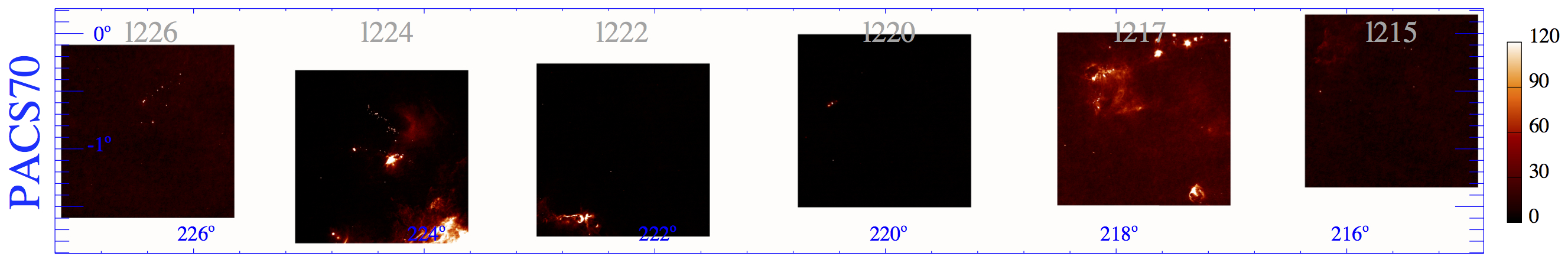} 
\includegraphics[width=17cm]{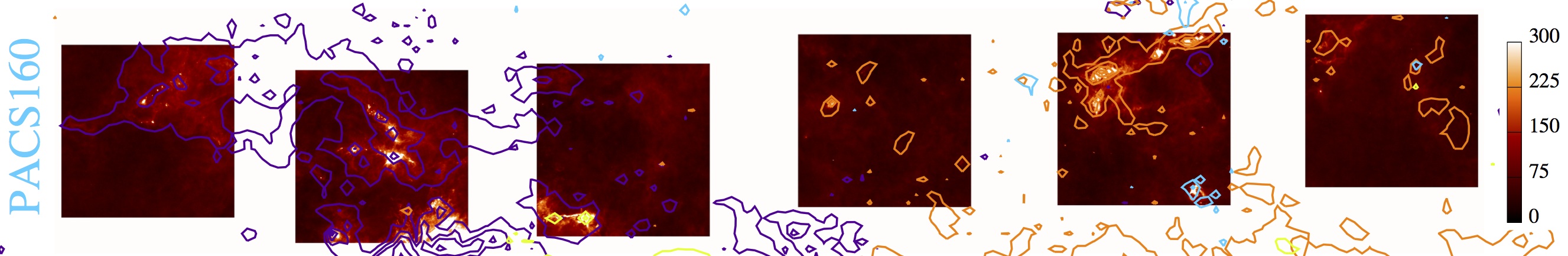} 
\includegraphics[width=17cm]{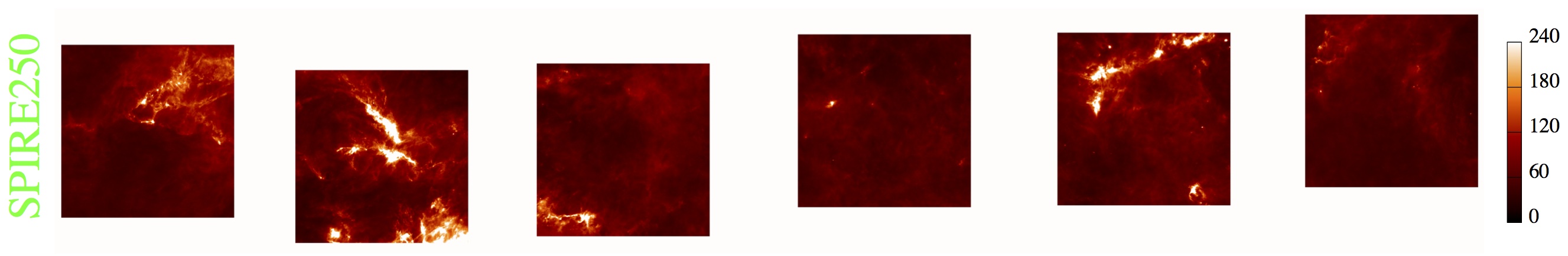} 
\includegraphics[width=17cm]{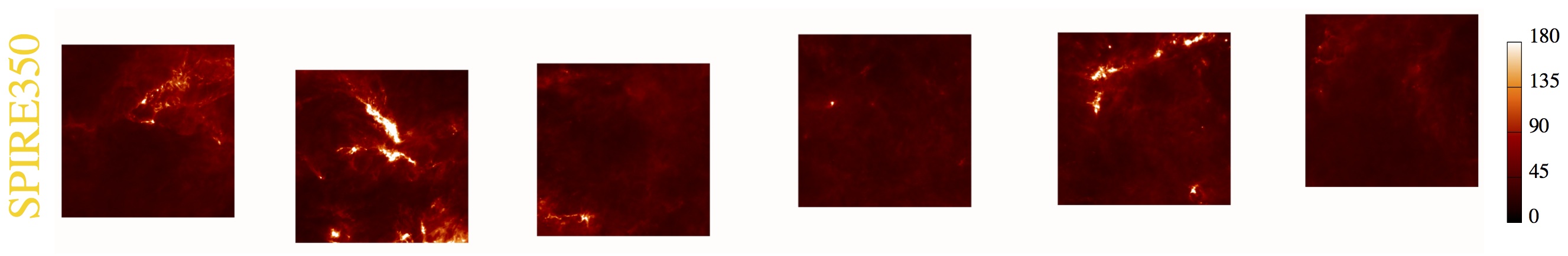} 
\includegraphics[width=17cm]{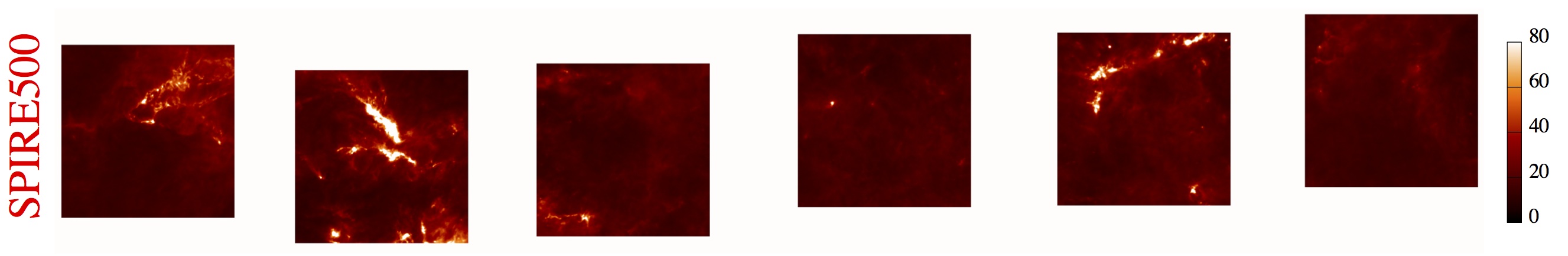} 
\includegraphics[width=17cm]{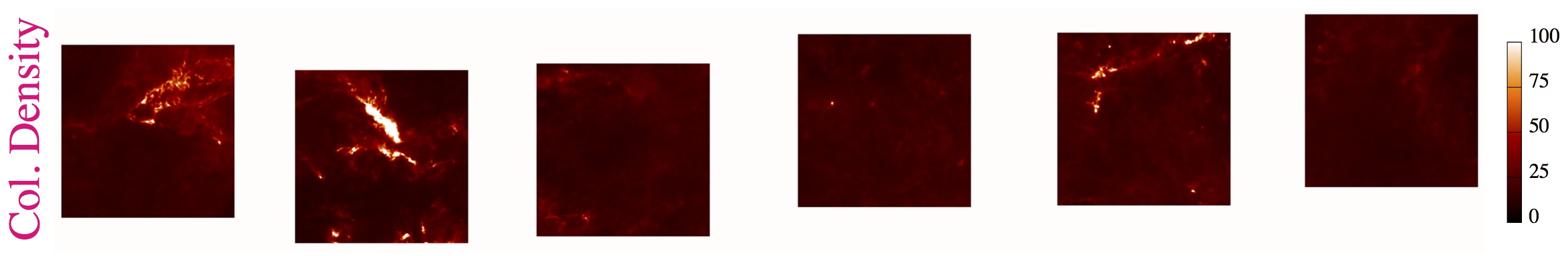} 
\caption{Each row shows the six Hi-GAL fields investigated in this paper. The instrument/band 
combination is specified in each row, adopting the color convention that is used throughout the 
rest of the paper, namely blue: 70~$\mu$m, cyan: 160~$\mu$m, green: 250~$\mu$m, yellow: 
350~$\mu$m, red: 500~$\mu$m, magenta: column density. The color scales are linear; we 
preferred this choice with respect to the logarithmic scale to show the actual dynamics of the 
maps processed in this paper.  The units are MJy~sr$^{-1}$ for the PACS and SPIRE maps, and 
$10^{20}$ cm$^{-2}$ for the column density ones. The tile names and Galactic coordinate grid 
are displayed in the row of 70~$\mu$m 
images (longitude on the horizontal axis and latitude on the vertical axis). Notice that the frame 
displacement is a real effect due to the choice of square area cut from each original tile.
The CO(1-0) contour levels of \citet{eli13} are overplotted on the row of 160~$\mu$m images. 
The contours start from 
5~K~km~s$^{-1}$ and are in steps of 15~K~km~s$^{-1}$. Velocity components I, II, III, and IV 
(see text) are represented with purple, orange, yellow, and cyan contours, respectively.}\label{htiles}
\end{figure*}

Furthermore, we include the two Hi-GAL tiles $\ell215$ and $\ell226$ that extend the sample to the 
West and to the East, respectively. In total, we have six tiles in five bands plus column density maps, 
i.e. a total of 36 maps analysed in this work. All maps are displayed in Figure~\ref{htiles}.

In \citet{eli13} an analysis of the gas velocity field, as derived from NANTEN CO(1-0) spectral line 
observations, was carried out for the four central tiles. It highlighted a prominence of emission from a 
closer component (at distances around $\langle d\rangle_\mathrm{I}=1.1$~kpc) for the two Eastern tiles $\ell224$ 
and $\ell222$, and from a farther component (around $\langle d\rangle_\mathrm{II}=2.2$~kpc) for the two Western 
ones, $\ell220$ and $\ell217$. These two components are also spatially segregated, since their 
reciprocal contamination degree is very low in the NANTEN data. Moreover, the contamination from two 
further and fainter velocity components ($\langle d\rangle_\mathrm{III}=3.3$~kpc, 
$\langle d\rangle_\mathrm{IV}=5.8$~kpc), is negligible in turn, even more after appropriate map 
cropping (see Figure~\ref{htiles}). In the four central panels of the 
second row in Figure~\ref{htiles} the low degree of contamination for the four corresponding
Hi-GAL fields can be appreciated. This made it possible, in \citetalias{eli14}, to associate 
each tile to only one of the two main velocity components of the region. 
In this paper, with the additional $\ell215$ and $\ell226$ tiles, we extend the basic 
analysis of the velocity field to these tiles. Looking at CO intensity contours obtained
separately for each of the four aforementioned Galactic components, and overplotted on Hi-GAL 160~$\mu$m 
maps (Figure~\ref{htiles}, second row, the leftmost and rightmost tile, respectively), one can see how 
the region in $\ell226$ is dominated by component~I,
whereas the one in $\ell215$ is dominated by component~II, so that, in conclusion, we can associate
the three Eastern tiles to component~I, and the three Western ones to component~II, respectively.

\subsection{Turbulent ISM simulations}\label{simulations}
\begin{figure}
\includegraphics[width=8.0cm]{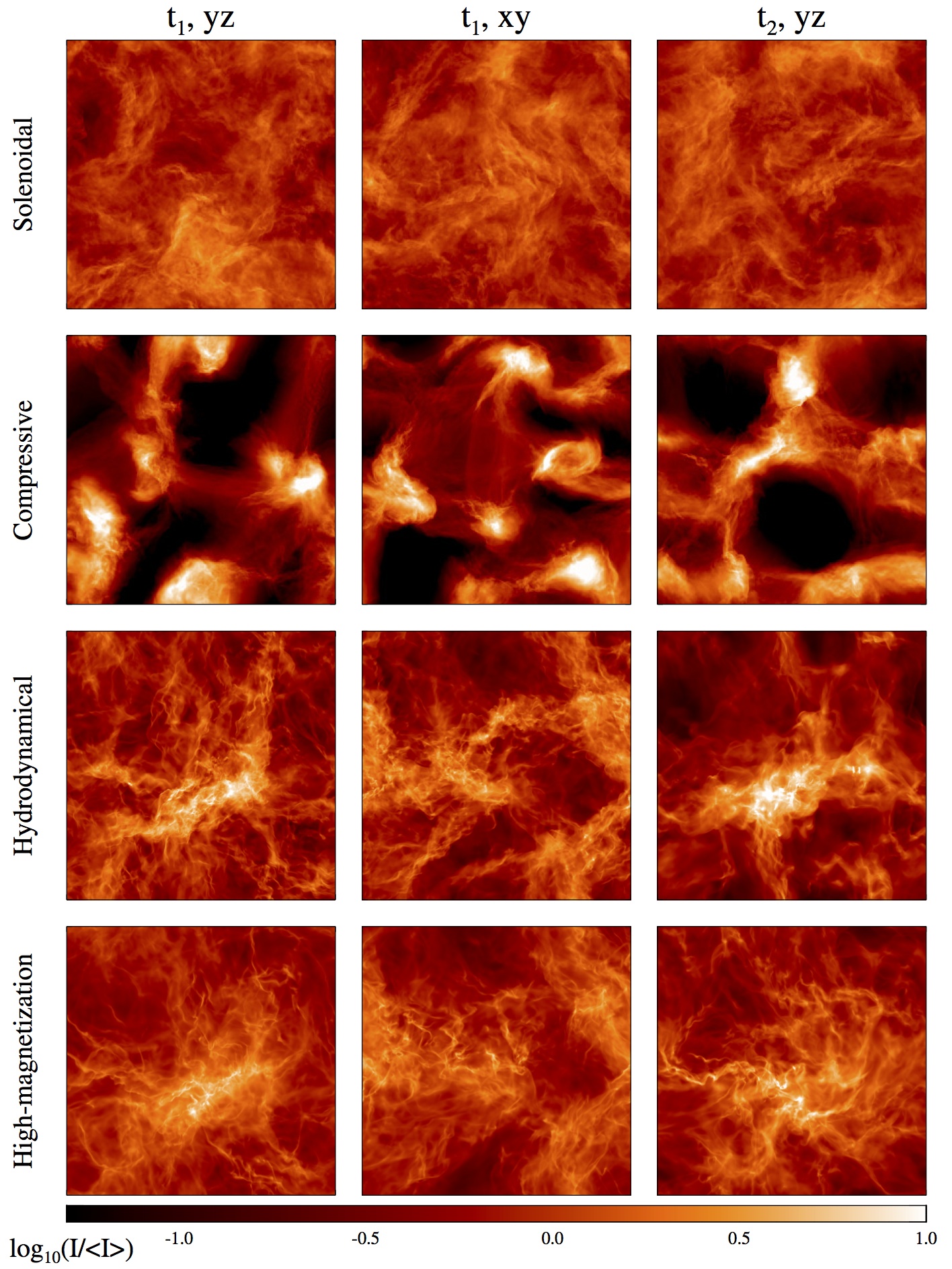} \caption{Column density maps obtained by
projecting 3-dimensional simulated density fields selected in the STARFORMAT archive.
The rows, from top to bottom, correspond to four different scenarios, namely ``solenoidal 
forcing'', ``compressive forcing'', ``quasi-hydrodynamical'', and ``high-magnetization'',
respectively (see text). The images in the upper two rows are composed by 
$256 \times 256$~pixels, while those in the lower two rows are composed by 
$2048 \times 2048$~pixels, see Table~\ref{imatab}. In each column a different 
combination of epoch and projection direction is displayed: the assignment of $yz$ 
and $xy$ plane denominations follows the notation used in STARFORMAT, 
used here only to indicate that the two planes are orthogonal. With $t_1$ 
and $t_2$ we do not indicate, in general, the same time for all the 
four scenarios, but only two different epochs such that $t_1 < t_2$; they coincide for the 
first and the second scenario only. Further details of the simulations are reported in the 
text. Maps are represented here by arbitrarily normalising their values by the map 
average $\langle I\rangle$, and rendered through a logarithmic scale, choosing an image 
saturation level of $\log_{10}(10 \langle I\rangle)$, to better appreciate the low-density fluctuations.}\label{modelimage}
\end{figure}

We compare the observational data to the density structures found in 3-dimensional
numerical simulations provided by the STARFORMAT project\footnote{starformat.obspm.fr}, 
a database containing results of high-resolution numerical simulations computed to study the 
formation, evolution and collapse of molecular clouds\footnote{Among other interesting resources 
for downloading numerical simulations of the ISM, we advide also Galactica 
(http://www.galactica-simulations.eu), CATS (http://mhdturbulence.com), and the
John's Hopkins Turbulence Databases (http://turbulence.pha.jhu.edu), not used for this 
work.}. Among the available projects, we chose the two we describe in the following. The 
``Solenoidal vs. compressive turbulence forcing'' project was carried out by \citet{fed08} in 
order to investigate differences between models of these two limiting regimes characterised by 
different mechanisms of kinetic energy injection. Simulations were performed using the 
FLASH3 code \citep{fry00,dub08}, solving hydrodynamic equations on $1024^3$ cubes, assuming
an isothermal gas (self-gravity is not included). Snapshots of the cube are available, for 
both scenarios, at different time steps,
calibrated on the timescale parameter $\mathcal{T}=\mathcal{L}/(2 c_s \mathcal{M})$, where $\mathcal{L}$
is the size of the computational domain, $c_s$ the sound speed and $\mathcal{M}$ the rms Mach number. 
To explore possible structure variations at two different epochs, we considered the second and the 
second last available time steps, namely for $t_1=3\mathcal{T}$ and $t_2=9\mathcal{T}$.

The used data sets consist of the column density maps obtained projecting the density cube along 
one of the three axes, say $x$ (then, in this convention, in the $yz$ plane), and rebinned onto a 
$256^2$ pixel grid. Moreover, to 
investigate possible projection effects, also the projection in the $xy$ plane of the cube at 
the time $t_1$ has been used. In conclusion, exploring two scenarios 
(solenoidal and compressive), two epochs and an alternative projection direction for the first 
epoch, six different maps were included in our analysis (Figure~\ref{modelimage}, top two rows). 
Images corresponding to the two different scenarios appear quite different, being the compressive forcing
maps characterised by a sharper contrast between very bright and extended void regions. Differences in 
structure have been already highlighted and quantified by \citet{fed08,fed09,fed10} through probability 
density function and power spectrum/$\Delta$-variance analysis. In \citetalias{eli14} a generally better
consistency in terms of power spectrum slope/fractal dimension has been found between the four central
Hi-GAL fields analysed in this paper and the solenoidal case rather than the compressive one.

The second project selected in the STARFORMAT data base is ``Molecular cloud evolution with 
decaying turbulence'' \citep{dib10,sol13}, which aims at describing a self-gravitating 
cloud with decaying turbulence in presence of a magnetic field. We considered two limiting 
scenarios with a weak and a strong mean magnetic field ($B<4~\mu \mathrm{G}$ and $B>20~\mu \mathrm{G}$,
respectively), called ``quasi-hydrodynamical'' and ``high-magnetization'' case, respectively. Simulations 
were performed through the RAMSES-MHD code \citep{tey02,fro06} and produced cubes with an effective 
resolution of $2048^3$ elements reproducing a 4 pc-wide region 
hosting a cloud with a mass of $\sim 2000~\mathrm{M}_\odot$. Two snapshots are 
available, in correspondence of $t_1=0.49$~Myr and $t_2=1.16$~Myr for the former scenario, and 
$t_1=0.61$~Myr and $t_2=1.15$~Myr for the latter, respectively. Also in this case, the projection of 
the density field in the $yz$~plane has been taken at these two epochs, together with the projection
in the $xy$~plane at the earliest of the two epochs. The selected fields are shown in the last
two rows of Figure~\ref{modelimage}.

\subsection{\textit{Fractional Brownian motion} images}\label{fbm} 
The \textit{fractional Brownian motion} images \citep[fBm,][]{pei88} are often used as a surrogate
of ISM maps thanks to their visual similarity with cloud features \citep{stu98,ben01,miv03}. Their 
analytic properties are fully described in \citet{stu98} and recently resumed in \citetalias{eli14}, 
to wich we refer the reader for more details. Here, we simply remind their basic properties: 
$i$) their radially averaged power spectrum exhibits a power-law 
behaviour, and $ii$) the distribution of the phases of their Fourier transform is completely random.
These images are fractal, with a precise analytic relation between their (mono-)fractal dimension 
$D$ (see Section~\ref{multi}) and the power-spectrum power-law exponent $\beta$:
\begin{equation}\label{dfbm}
D=4-\frac{\beta}{2}\;,
\end{equation} 
which is valid for a signal defined over a two-dimensional space \citep{stu98}.

In this paper, we use these images as a reference, since they can be obtained with preconditioned
statistical properties. In particular we generated a set of $1000\times 1000$ pixels fBm images 
by exploring a two-dimensional parameter space, i.e. varying both the $\beta$ exponent (then the 
global fractal dimension) from 2 to 4 in steps of 0.4 \citepalias[cf.][]{eli14}, and the distribution 
of the phases, initializing the random number generator with three different ``seeds'', called hereafter 
realization A, B, and C. A given phase distribution determines the global appearance of the image 
as a ``cloud'', while increasing $\beta$ produces a gradual smoothing of the image, due to the transfer 
of power from high to low spatial frequencies, as it can be seen in Figure~\ref{fbmfig}  
(Appendix~\ref{fbmappend}).

\section{Multifractal analysis}\label{multi}
As pointed out in Section~\ref{intro}, the multifractal analysis represents an extension 
of the fractal theory, aimed at enhancing and enriching the characterization of a 
non-deterministic (i.e. natural) fractal. It is closely related to the concept of 
\emph{generalized fractal dimension}.
%, that will be introduced as preliminary remarks.

\subsection{Generalized fractal dimension}\label{multidim}
The \textit{fractal dimension}, introduced by \citet{man67}, is one of the fundamental concepts
in fractal analysis. It expresses the degree of complexity of a self-similar object, and in 
particular its ability to fill the hosting space. Importantly, it is not an integer number, 
exceeding the Euclidean dimension of the set 
(e.g., the fractal dimension of a fractal line is larger than 1, the one of a fractal surface 
is larger than 2, etc.). In this respect, the fractal dimension represents a meaningful indicator
for quantifying the structure of complex, nested, convoluted structures which depart from the 
smooth appearance of Euclidean shapes. Nevertheless, the need of going beyond a single 
descriptor, leads to the formulation of the multifractal geometry, based on the concept of 
\emph{fractal generalized dimension}.

The definition of a set of dimensions of order $q \in \mathds{R}$ is due to \citet{hen83}, 
who introduced this concept to improve the characterization of chaotic attractors 
of dynamical systems. 
% In such cases, the fractal dimension might be hard to calculate, since
%rarely visited regions of the attractor might contribute significantly to it. Furthermore, 
A single parameter constitutes an incomplete characterization of such sets, which are not 
perfectly self-similar by construction like, instead, the deterministic 
examples (such as the well-known Cantor set or the Koch curve). The \emph{generalized dimensions}, 
therefore, represent an attempt to address in a more general way the characterization of the 
so-called ``strange sets''.

Among various formulations of the generalized dimensions, here we briefly report the one based 
on the box-counting approach \citep[cf.][]{hen83}: let us start considering a set constituted by 
points (as, for example, black pixels in an image with a white background), in this
context generally named \textit{measure}, and cover it with boxes 
of size $\varepsilon$. Let us define the probability of points contained in the $i$-th box as 
$P_i = N_i/N$, where $N_i$ is the number of points in the $i$-th box, out of a total number of 
points $N$. The $q$-th generalized dimension $D_q$ is defined as
\begin{equation}\label{generalbox}
D_q =\lim_{\varepsilon\rightarrow 0}\frac{1}{q-1}\frac{\log\sum_i
P_i^q}{\log\varepsilon}\; .
\end{equation}

Defining the partition function of the $q$-th order as
\begin{equation}\label{zqfunction}
Z_q(\varepsilon)= \sum_i P_i^q\;,
\end{equation}
the definition in Equation~\ref{generalbox} becomes
\begin{equation}\label{general_z}
D_q =\lim_{\varepsilon\rightarrow 0}\frac{1}{q-1}\frac{\log
Z_q(\varepsilon)}{\log\varepsilon}\;.
\end{equation}

Notice that the case $q=0$ recovers the classic expression of the box counting mono-fractal
dimension $D$ \citep[e.g.,][]{fal03}. An interesting case is $D_1$ (also called 
\emph{information dimension}),
obtained from Equation~\ref{generalbox} in the limit $q\rightarrow 1$. By construction, 
all the $D_q$ dimensions of a deterministic pure fractal will coincide with $D$, while a 
\emph{multifractal} is an object whose generalized dimensions are quite different. 
The $D_q$ values monotonically decrease with increasing $q$ \citep{hen83}, and the lower and 
upper limiting dimensions, named $D_{-\infty}$ and $D_{+\infty}$, respectively, are related to 
the regions of the set in which the measure is ``most diluted'' and ``most concentrated'',
respectively.

In general, the partition function of multifractal sets scales as
\begin{equation}\label{zetaq}
Z_q(\varepsilon) \propto \varepsilon^{\tau_q}\; ,
\end{equation}
\citep[e.g.,][]{arn95}, with $\tau_q$ named \emph{correlation exponent} of order $q$. The 
relationship between $\tau_q$ and $q$ is expressed by
\begin{equation}\label{tauh}
\tau_q=q\;h(q)-E\;,
\end{equation}
where $E$ indicates the Euclidean dimension of the volume hosting the fractal, named
\textit{support of the measure} \citep[e.g.,][]{gu06}.
For a monofractal, the $h(q)$ coefficient is constant (and called \textit{Hurst exponent}) so 
that the $\tau_q$ vs $q$ relation is linear, whereas in the more general case of a multifractal 
$h(q)$ (in such case called \emph{generalized Hurst exponent}) varies as a function of $q$. 

Finally, combining Equations~\ref{general_z} and \ref{zetaq} one finds
\begin{equation}\label{Dq}
\tau_q=(q-1)D_q
\end{equation}
\citep{hal86}. As a particular case of this equation, it is found that $D_0=-\tau_0$.

At this point, some caveats about the interpretation of the introduced quantities and their
application to astronomical maps (or to grey-scale images in general) need to be pointed out. Indeed, 
all the above definitions and considerations are strictly valid for a measure composed 
by discrete points, when a coverage of the support of the measure is performed. In our 
case, the measure is a 3-D surface having values different from zero at any point of its 
support, which is a $m\times n$ pixels raster. The probability $P_i$ is the sum of the brightnesses
of all pixels falling inside the $i$-th box, normalized by the integral over the entire map.
With this approach, it can be seen that, for $q=0$, Equation~\ref{generalbox} does 
not return the true fractal dimension of the investigated image, but rather the Euclidean 
dimension of the support of the measure, in other words one finds $D_0=2$ for all images.

%\begin{figure}
%\includegraphics{dqvela.eps} 
%\caption{.}
%\label{dqvela}
%\end{figure}
%
%\begin{figure*}[p]
%\includegraphics{dqfbm.eps} \caption{.}\label{dqfbm}
%\end{figure*}
%
%In Figure \ref{dqvela} the $\tau_q$ and $D_q$ curves for one of the
%test fBm images are displayed, while in Figure \ref{dqfbm} we do the
%same for THE INTEGRATED INTENSITY MAP !!!! No, almeno due fBm!!! The
%$q$ parameter spans the range [-15,20], in steps of 1, except the
%subrange [-2,1] that is a more critical region and then was covered
%with steps of 0.2. A first look already reveals that both the
%considered maps show an evident multifractal-like behavior, although
%it is more evident for the VMR-D map, that therefore
%XXXXXXXXXXXXXXXXXXXXX. Furthermore, in the region of positive $q$
%values the curves of the fBm images appear to be quite similar, as we
%will also see in the following. Finally, note that in all cases we
%find $D_0=2$, as expected.
%
%A more practical
%representation of the multifractal properties of an investigated set,
%namely the \emph{multifractal spectrum}, is described in next section.

\subsection{Multifractal spectrum}

A tool typically used for the multifractal analysis is the so-called \emph{singularity spectrum}, or
\emph{multifractal spectrum} (MFS), first introduced by \citet{hal86}.

Let us consider a measure (here our Hi-GAL maps) embedded in a support and cover it with boxes 
of size $\varepsilon$ as already discussed in Section \ref{multidim}. If the measure is a multifractal,
the probability $P_i$ scales with $\varepsilon$ as a power law with exponent $\alpha_i$ (also 
known as \textit{singularity strength}), as a function of the position: $P_i(\varepsilon)\propto
\varepsilon^{\alpha_i}$. Given the fractal dimension $f(\alpha)$ of the subset of boxes having 
singularity strength in the range $(\alpha,\alpha+d\alpha)$, the MFS is just the curve of $f(\alpha)$
vs $\alpha$. It represents the contributions to the geometry provided by interwoven sets 
with different singularity strengths.

The relation between the MFS and the generalized dimensions is expressed by the
\emph{Legendre's transform}:
\begin{equation}\label{legendrea}
\alpha(q) = \frac{d\tau_q}{dq}
\end{equation}
\begin{equation}\label{legendref}
f(\alpha(q)) = q\alpha(q)-\tau_q 
\end{equation}
\citep{hal86}. In this sense, the MFS can be calculated immediately after deriving the ($q,D_q$) pairs.

From these relations the following properties of the $f(\alpha)$ curve are derived:
\begin{equation}
\frac{df}{d\alpha}=q
\end{equation}
\begin{equation}
\frac{d^2f}{d\alpha^2}<0\,.
\end{equation}
Therefore, the MFS is concave for any measure, with a single maximum at $q=0$ (then $f(\alpha(0))=D_0$
for a measure constituted by points\footnote{As already pointed out in Section~\ref{multidim}, 
here wee analyse greyscale images. In this case, $f(\alpha(0))$ coincides with the 
Euclidean dimension of the support of the measure, namely 2, as it can be seen, in the following, 
from Equation~\ref{fq} combined with Equation~\ref{muq}.}), and with infinite slope at $q=\pm\infty$. 
%Moreover, it is to notice that $\alpha(1)=f(1)$.

The MFS gives information about the relative importance of various fractal exponents present in 
the map, and in particular its width indicates the range in which such exponents lie. The part
corresponding to values $q>0$ characterises the scaling properties of overdense regions because
it magnifies the effect of large numbers, while for $q<0$ the behaviour of low-density subsets is
characterised. This generally results in an asymmetrical shape of the MFS with respect to its peak 
position ($q=0$).

\subsection{Practical derivation of multifractal parameters}\label{mfscalc}

\begin{figure*}
\includegraphics[width=17cm]{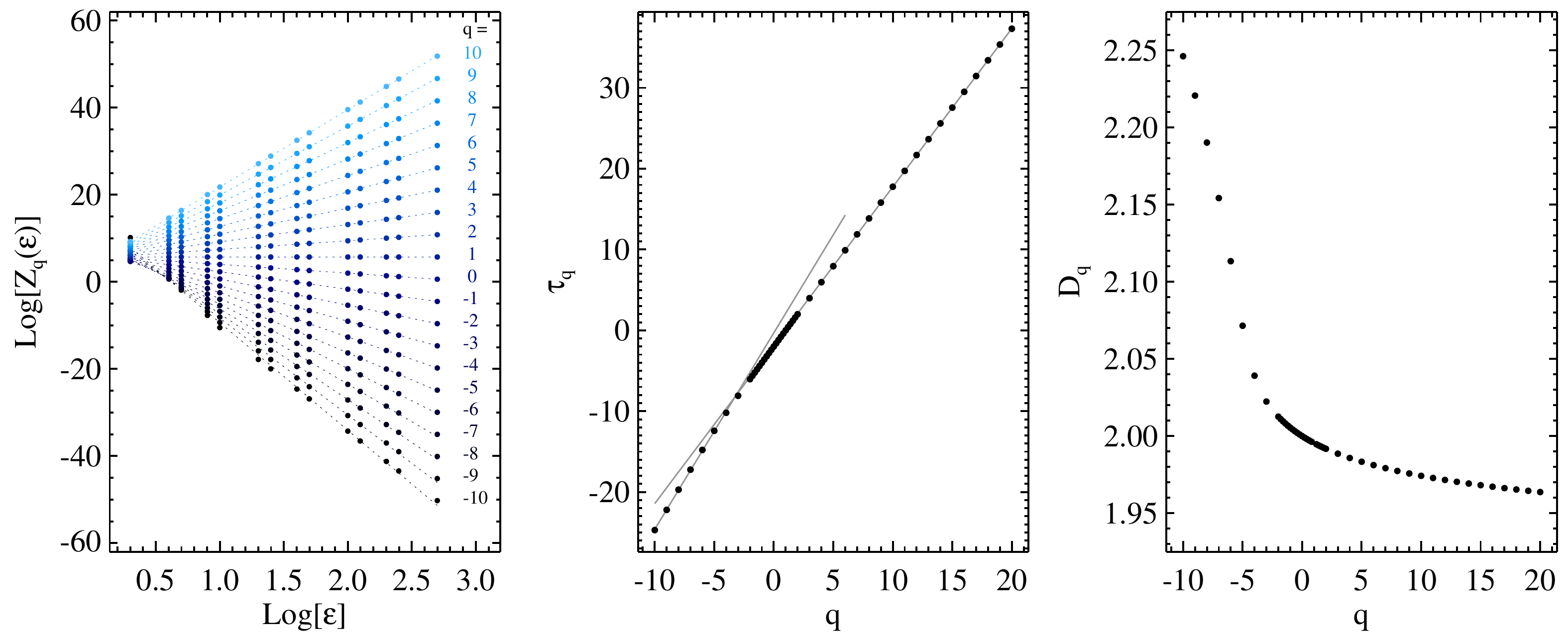} 
\caption{Example of derivation of the generalised fractal dimensions for a test 
$1000 \times 1000$-pixel fBm image (see Section~\ref{fbm}), in this case the one of type ``C'' 
and $\beta=2.4$ shown in Figure~\ref{fbmfig}. \textit{Left}: logarithm of the partition functions 
$Z_q(\varepsilon)$ vs the logarithm of the integer scales $\varepsilon$ between 1 and 500 pixels 
for each integer value of the parameter $q$ between -10 and~10 (in this range, different shades of 
blue, from the darkest to the lightest, respectively, are used for each different $q$). The 
linear fit is shown as a dashed line of the same colour of the 
corresponding $Z_q$ function (cf. Equation~\ref{zetaq}). The functions for orders $q>10$ and non-integer 
orders in the range $-2 < q < 2$, although computed, are not shown here for the sake of clarity. 
\textit{Middle}: scaling exponents $\tau_q$ of the partition function, obtained through the linear fit 
shown in the left panel as a function of $q$. To highlight the different slopes at negative
and positive orders, the linear fit of the $\tau_q$ vs $q$ in the two the regions $q<-3$ and $q>3$, 
is shown by means of two grey lines. \textit{Right}: generalised 
fractal dimension set, derived through Equation~\ref{Dq}, as a function of $q$, except for
the case $q=1$, not contemplated by Equation~\ref{Dq}.}
\label{allq}
\end{figure*}

For the purpose of this paper we derive the values of $\tau_q$ and $D_q$ for each map through the 
partition function, as described in Section~\ref{multidim}. The analysed sets are constituted
by a discrete number of pixels, thus we partition the data sets in boxes with integer size $\varepsilon$.
In this respect, it is preferable to deal with image sizes that have a large number of integer 
divisors (see Section~\ref{higal}) enough to ensure a sufficiently large range of investigated spatial 
scales, i.e. points to sample the behavior of the right-hand side of Equation~\ref{general_z}. 

In Figure~\ref{allq} we show a derivation of these quantities for a fBm test image 
%which is described in Section~\ref{fbm} and Appendix~\ref{fbmappend}, and 
shown in Figure~\ref{fbmfig}, third-row and third-column panel, corresponding to 
realization ``C'' and power spectrum slope 2.4. 
For this example, as well as for all maps analysed in this paper, the probed range of orders 
$q$ is $-10 \leq q \leq 20$, in steps of $\Delta q=1$, except for the interval $-2 \leq q \leq 2$, 
which is explored in steps of $\Delta q=0.2$ to better sample a critical region
for multifractal descriptors (slope change for $\tau_q$, inflection point of $D_q$, and peak of 
the MFS around $q=0$, see below).

\begin{figure}
\includegraphics[width=8.5cm]{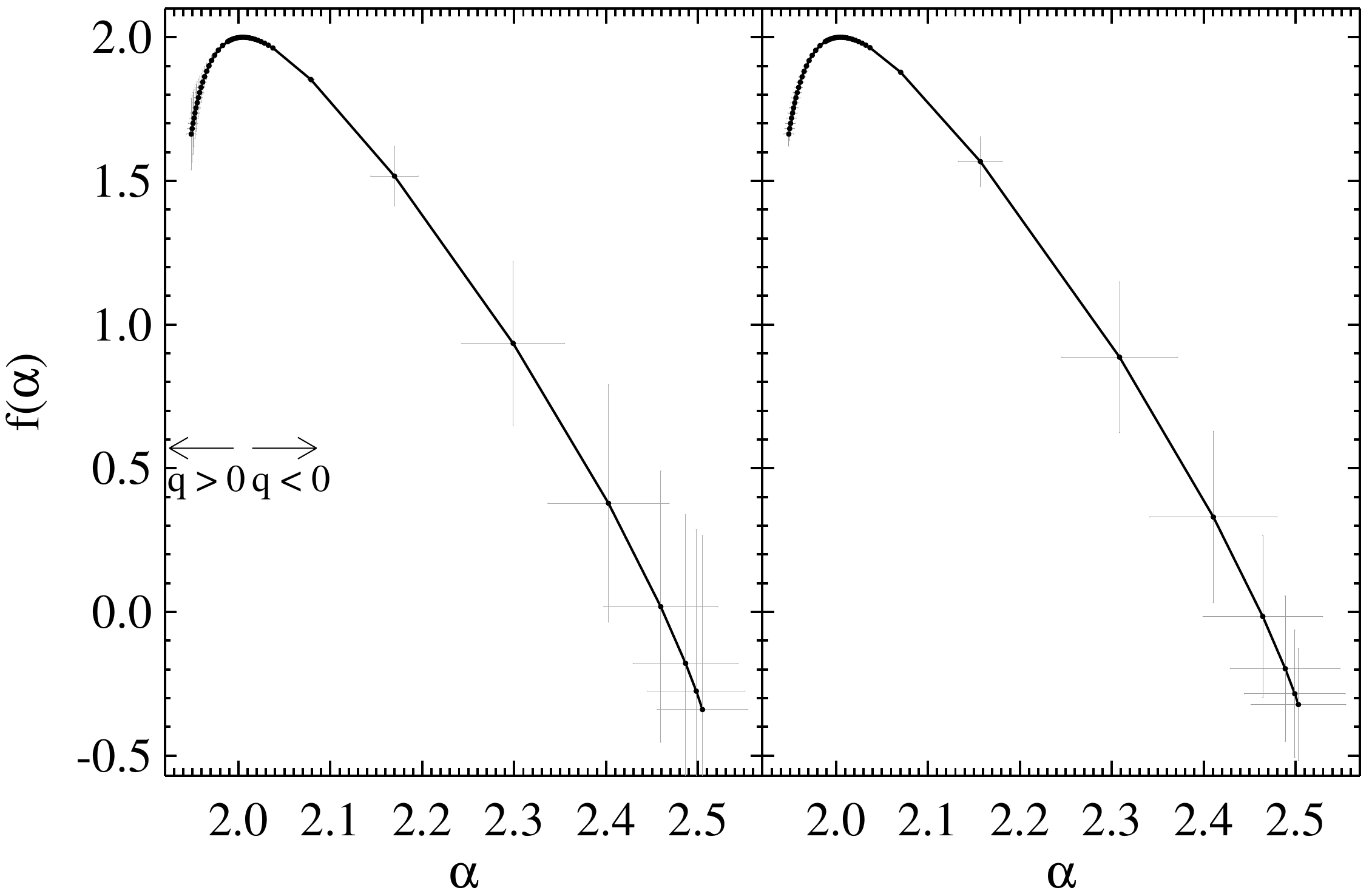} 
\caption{\textit{Left}: MFS of the same fBm reference image used to build Figure~\ref{allq}, 
obtained through Equations~\ref{legendrea} and ~\ref{legendref}, exploring the range of orders 
$-10 \leq q \leq 20$, in steps of 
$\Delta q=1$ (but $\Delta q=0.2$ in $-2 \leq q \leq 2$). The error bars are displayed as grey crosses
for each point, to give an idea of the typical uncertainties affecting the MFS estimation.
\textit{Right}: The same as in the left panel, but computed through Equations~\ref{aq} and~\ref{fq}.}
\label{fbmmfs}
\end{figure}

The $Z_q(\varepsilon)$ function (left panel) shows the expected power-law behaviour, 
from which the $\tau_q$ scaling fits can be derived through a linear regression procedure 
(Equation~\ref{zetaq}). The quality of such fit determines 
in turn the extent of the error bar on $\tau_q$. The $\tau_q$ vs $q$ curve (middle panel) does not 
present a single linear behaviour, but a typically multifractal behaviour with two different 
slopes for the two $q<0$ and $q>0$ regimes, and a non-linear behaviour in the transition zone 
\citep[cf., e.g.,][]{moh11,xie15}. Similarly, the obtained $D_q$ vs $q$ curve (right panel) is 
far from being constant \citep[cf., e.g.,][]{hal86,men87}, confirming that the analysed image 
presents a multifractal rather than a simple mono-fractal character.

Once $\tau_q$ and $D_q$ are obtained, the MFS can be derived through Equations~\ref{legendrea}
and~\ref{legendref}.
Another linear fit is requested in this case, introducing a larger uncertainty on both $\alpha$
and $f(\alpha)$. An alternative method for calculating directly the MFS was developed by 
\citet{chh89} and applied by \citet{cha01} for estimating the MFS of \textit{IRAS} maps.
Defining the coarse-grained moments $\mu_i(q,\varepsilon)$ of the original image as
\begin{equation}\label{muq}
\mu_i(q,\varepsilon)=\frac{P_i^q(\varepsilon)}{\sum_i P_i^q(\varepsilon)}\,,
\end{equation}
$\alpha$ and $f(\alpha)$ are implicitly defined with respect to the $q$ parameter through
\begin{equation}\label{aq}
\alpha(q)=\lim_{\varepsilon\rightarrow 0}\frac{\sum_i \mu_i(q,\varepsilon) \log P_i(\varepsilon)}{\log \varepsilon}
\end{equation}
\begin{equation}\label{fq}
f(q)= \lim_{\varepsilon\rightarrow 0}\frac{\sum_i \mu_i(q,\varepsilon) \log \mu_i(q,\varepsilon)}{\log
\varepsilon} \,.
\end{equation}
In practice, in the case of an ideal multifractal, the quantities whose limits are calculated in 
the right sides of the equations above show a linear behavior, since the limits can be extrapolated 
from a linear fit.
Consequently, the uncertainties associated to $\alpha(q)$ and $f(\alpha(q))$ mainly depend on the 
quality of such fit. Again, we reaffirm, also for this approach, the need of analysing images with 
sizes divisible by a large number of integer factors.

In Figure~\ref{fbmmfs}, we compare the MFS obtained for the same fBm image using the two methods. The 
spectra look practically identical, with differences smaller than 1\% in all cases, while the error
bars, which are generally much larger for $q<0$, are comparable for $\alpha$ but larger for the first
metod for $f$. For this reason, in the following we will show MFSs obtained through the \citet{chh89} 
method (Equations~\ref{aq} and~\ref{fq}), keeping the first one as a reference for checking 
the correctness of the results.

\section{Results of multifractal analysis}\label{results}

The multifractal behaviour of the analysed maps can be examined by means of both the generalised 
dimension formalism ($D_q$ vs $q$) or the MFS one ($f(\alpha)$ vs $\alpha$). We start from the 
latter, which allows us to better highlight some peculiarities in our map sample with respect to the 
scaling implied by Equation~\ref{zqfunction}.

\subsection{Qualitative analysis of MFS}

\begin{figure*}
\includegraphics[width=17cm]{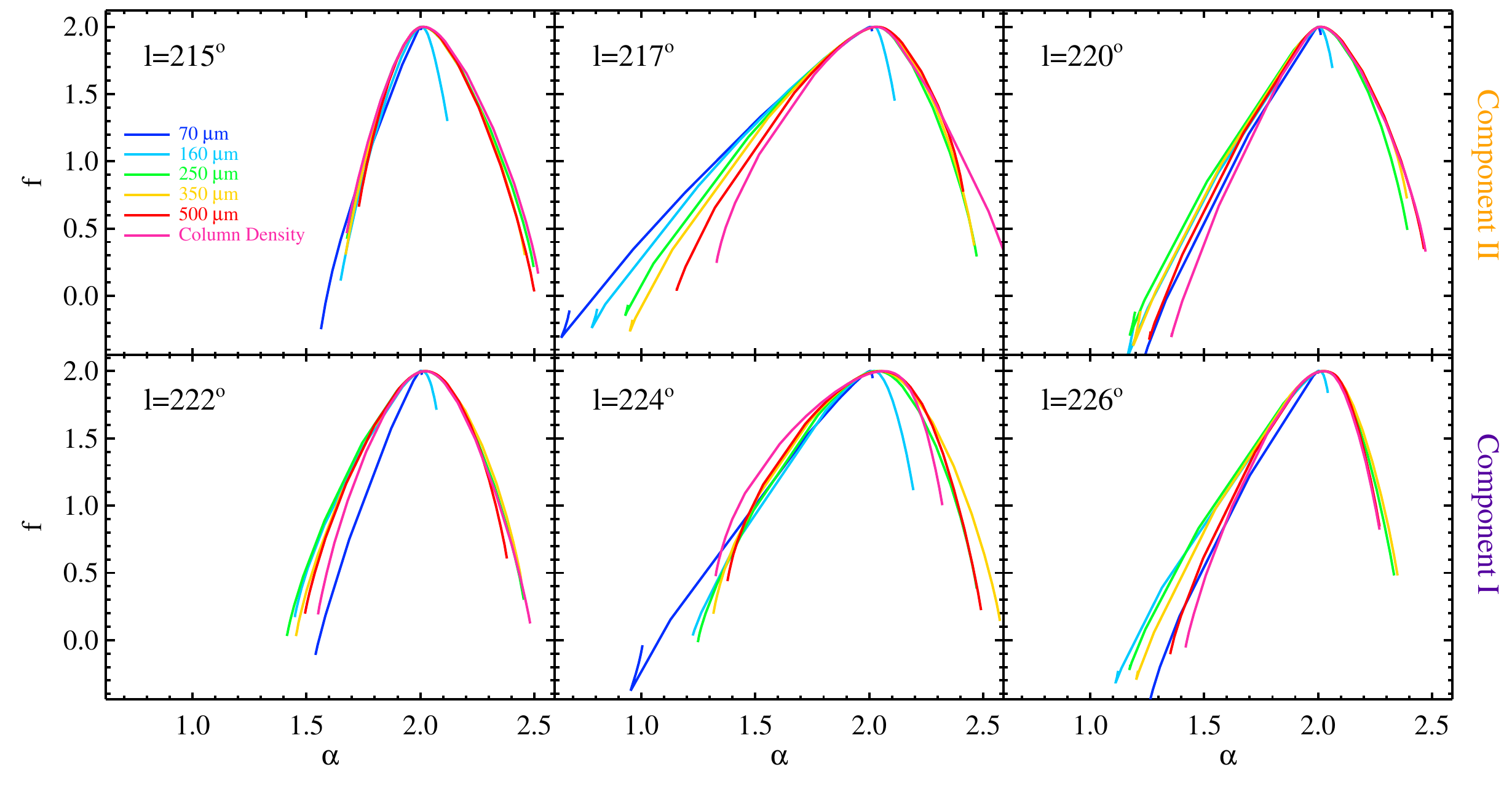} \caption{MFS of the Hi-GAL maps analysed 
in this work, ordered by tile (in turn, tiles associate to kinematic components I and II are
arranged in the lower and in the upper row, respectively). The colour-band correspondence is 
the same introduced in Figure~\ref{htiles} and used throughout the entire paper.}\label{mfsspectra}
\end{figure*}

\begin{table*}
\caption{Main properties of the MFS of the analysed Hi-GAL frames.}
\label{mfstab}
\begin{tabular}{llcccccccccc}
\hline
Field & Band & $\alpha_{20}$ & $f_{20}$ & $\alpha_{10}$ & $f_4$ & $\alpha_0$ & $\alpha_{-10}$ &  $q^+_c$ & $A_{\Phi}$\\
\hline
$\ell$215 & 70 &  1.53 & -0.64 &  1.56 &  1.72 &  2.001 &  2.01 & ... & ... \\
 & 160 &  1.63 & -0.06 &  1.65 &  1.77 &  2.006 &  2.12 & ... & 0.07 \\
 & 250 &  1.65 &  0.05 &  1.68 &  1.70 &  2.012 &  2.50 & ... & 0.12 \\
 & 350 &  1.64 & -0.06 &  1.67 &  1.73 &  2.012 &  2.46 & ... & 0.12 \\
 & 500 &  1.70 &  0.22 &  1.73 &  1.76 &  2.014 &  2.50 & ... & 0.12 \\
 & Column Density &  1.66 &  0.23 &  1.68 &  1.71 &  2.014 &  2.52 & ... & 0.13 \\
$\ell$217 & 70 & ... & ... & ... & -0.31 &  2.007 &  2.01 &  4 & ... \\
 & 160 & ... & ... & ... & -0.24 &  2.023 &  2.11 &  4 & 0.14 \\
 & 250 & ... & ... & ... & -0.12 &  2.031 &  2.47 &  5 & 0.18 \\
 & 350 & ... & ... & ... & -0.17 &  2.035 &  2.46 &  5 & 0.19 \\
 & 500 & ... & ... & ... &  0.22 &  2.035 &  2.41 &  6 & 0.19 \\
 & Column Density &  1.32 &  0.16 &  1.33 &  0.69 &  2.030 &  2.69 & ... & 0.17 \\
$\ell$220 & 70 & ... & ... &  1.21 & -0.03 &  2.002 &  2.01 & 10 & ... \\
 & 160 & ... & ... & ... & -0.09 &  2.006 &  2.06 &  6 & ... \\
 & 250 & ... & ... & ... & -0.03 &  2.012 &  2.39 &  5 & 0.14 \\
 & 350 & ... & ... & ... &  0.03 &  2.013 &  2.39 &  6 & 0.14 \\
 & 500 & ... & ... & ... &  0.31 &  2.013 &  2.46 &  7 & 0.14 \\
 & Column Density & ... & ... & ... &  0.67 &  2.012 &  2.47 &  8 & 0.13 \\
$\ell$222 & 70 & ... & ... & ... &  1.57 &  2.002 &  2.01 &  9 & ... \\
 & 160 &  1.44 &  0.07 &  1.45 &  0.85 &  2.010 &  2.07 & ... & ... \\
 & 250 &  1.41 &  0.00 &  1.41 &  0.89 &  2.020 &  2.45 & ... & 0.15 \\
 & 350 &  1.45 & -0.06 &  1.45 &  1.08 &  2.021 &  2.44 & ... & 0.15 \\
 & 500 &  1.49 &  0.13 &  1.49 &  1.17 &  2.024 &  2.38 & ... & 0.15 \\
 & Column Density &  1.54 &  0.07 &  1.55 &  1.40 &  2.021 &  2.48 & ... & 0.14 \\
$\ell$224 & 70 & ... & ... & ... & -0.38 &  2.005 &  2.02 &  4 & ... \\
 & 160 &  1.23 &  0.04 &  1.23 &  0.20 &  2.022 &  2.19 & ... & 0.13 \\
 & 250 &  1.24 & -0.07 &  1.25 &  0.42 &  2.035 &  2.47 & ... & 0.18 \\
 & 350 &  1.31 &  0.07 &  1.32 &  0.71 &  2.044 &  2.57 & ... & 0.20 \\
 & 500 &  1.36 &  0.22 &  1.38 &  0.88 &  2.053 &  2.49 & ... & 0.21 \\
 & Column Density &  1.31 &  0.31 &  1.33 &  0.90 &  2.064 &  2.32 & ... & 0.20 \\
$\ell$226 & 70 & ... & ... &  1.25 &  0.18 &  2.001 &  2.00 & 12 & ... \\
 & 160 & ... & ... & ... & -0.21 &  2.007 &  2.04 &  5 & ... \\
 & 250 & ... & ... & ... &  0.08 &  2.021 &  2.33 &  6 & 0.14 \\
 & 350 & ... & ... & ... &  0.06 &  2.025 &  2.35 &  6 & 0.14 \\
 & 500 & ... & ... &  1.35 &  0.61 &  2.026 &  2.27 & 10 & 0.14 \\
 & Column Density & ... & ... &  1.42 &  1.00 &  2.024 &  2.27 & 11 & 0.13 \\
\hline
\end{tabular}
\end{table*}

In Figure~\ref{mfsspectra} the MFS of each Hi-GAL tile at each band are shown, while in Table~\ref{mfstab}
a number of salient parameters of the MFS, some of which introduced in the following 
Section~\ref{quantifying}, are listed\footnote{\label{simpnote}Notice that hereafter, for the MFS coordinates 
corresponding to a given order $Q$ we use the simplified notation $\alpha_Q$ and $f_Q$ in place 
of $\alpha(q=Q)$ and $f(q=Q)$, respectively.}. It is possible to make some initial and general 
considerations about the overall appearance of the spectra:

\begin{itemize}
\item In general, the MFSs present a typical single-humped shape with a maximum at $f_0=2$ (i.e. the 
Euclidean dimension of the support, as explained previously), but a strong negative skewness.
%, with the portion of spectrum left of peak (corresponding to $q>0$) generally more elongated 
%than the one after the peak ($q<0$).
\item In several cases, at high $q$ (i.e. at the left end of the MFS) a cusp-like turnback
of $\alpha(q)$ appears in the spectrum (the most evident case being found for the 70~$\mu$m map of 
$\ell224$, but other cases are well recognisable in the plots of Figure~\ref{mfsspectra}). Such a 
feature \citep[cf.][]{lou15} can be considered $i$) as a computational artefact, due to the presence of 
spatial heterogeneity of brightness in the image which leads to disproportionate contribution of boxes
containing rare extremely intense pixels during averaging over the whole area, and $ii$) as a real 
effect related to intrinsic statistical properties of the analysed image departing from those 
of a multifractally distributed measure. \citet{wol89}, indeed, demonstrated that this feature appears 
in correspondence of a breakdown in the scaling law expressed by Equation~\ref{zetaq} at orders
higher than a critical value $q^+_c$ \citep[see also, e.g.,][]{abr04}. Most likely, the appearance 
of such a cusp is produced by a combination of these two effects in images presenting a wide dynamic 
range between the diffuse emission and few very bright compact sources. In Table~\ref{mfstab} a 
possible presence of a cusp in the MFS is reported by
listing the possible value of $q^+_c$. The maps affected by this problem will suffer of some
limitations throughout the analysis carried out in this paper, since the part of the spectrum 
corresponding to $q>q^+_c$ (missing data in Table~\ref{mfstab}) will not be taken into account.
\item  In most cases, the right portion of the MFS for the 70~$\mu$m maps turns out to be collapsed
to a very short branch close to the peak, while the left portion (in correspondence
of positive $q$ values) appears extremely elongated. In general, a MFS shows a long left tail when 
the signal has a multifractal 
structure which is insensitive to the local fluctuations with small amplitudes, and a shrunk right 
tail in presence of a small scale noise-like background \citep[e.g.,][]{dro15}. This is the typical 
case of Hi-GAL 70~$\mu$m maps in this portion of the Galaxy generally showing rare and isolated
bright spots, corresponding to episodes of star formation, and a very faint background emission
often surmounted by the instrumental noise. This fact, together with the frequent occurrence,
for the 70~$\mu$m maps, of the aforementioned cusps in the MFS, makes the maps at this wavelength 
a peculiar case in our multi-fractal analysis, showing in general a departure from a fractal behaviour.   
\item For each tile, the MFSs of the SPIRE images show a relatively similar behaviour, but not identical.
The positions of the extreme points can differ, even significantly, from band to band depending on the
tile, the strongest differences being found in the tiles containing the brightest features, i.e.
$\ell217$, $\ell224$, and $\ell226$. Such overall similarities and fine-scale differences will be 
highlighted in the following more quantitative analysis.
\item The 160~$\mu$m maps show a behaviour of the MFS width which is intermediate between the 
70~$\mu$m and the SPIRE ones, being the $q<0$ tail less pronounced than for the SPIRE maps, but 
not collapsed as generally found for the 70~$\mu$m ones.
\item Finally, the column density maps, which could be expected to have a behaviour similar to the SPIRE
ones, actually differ significantly from them in some cases, especially for the tiles $\ell217$ and 
$\ell224$.
\end{itemize}

\begin{figure}
\includegraphics[width=8.5cm]{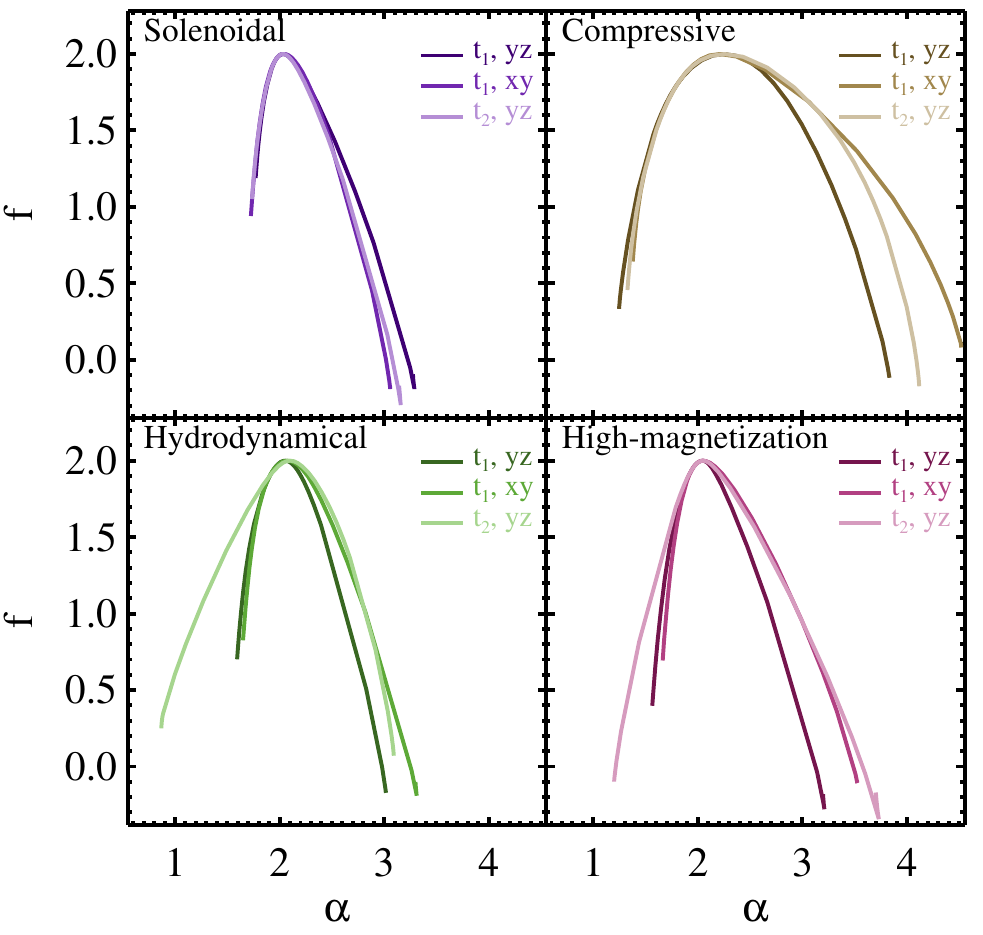} \caption{MFS of the column density maps obtained 
from simulations, shown in Figure~\ref{modelimage}. In each panel the model, the projection 
plane and the epoch of the simulation are specified for each displayed MFS. The range of 
$\alpha$ on the horizontal axis is chosen to optimise the plot of the represented MFS, and for
this reason a visual comparison with the plots in Figure~\ref{mfsspectra} (set according 
to the same criterion) can not be performed directly.}\label{modelmfs}
\end{figure}

The MFS was derived also for all the simulated column density fields described in 
Section~\ref{simulations} and shown in Figure~\ref{modelimage}. All the computed
MFSs are plotted in Figure~\ref{modelmfs}, each panel corresponding to a different
model. In the following, we summarize some results of the comparison among the 
different models and their projections and epochs, and with the Hi-GAL  observations 

\begin{table*}
\caption{Main properties of the MFS of the analysed simulations.}
\label{modmfstab}
\begin{tabular}{llcccccccccccccc}
\hline
\# & Model & Plane & Epoch & $\alpha_{20}$ & $f_{20}$ & $\alpha_{10}$ & $f_{4}$ & $\alpha_{0}$ & $\alpha_{-10}$ &  $q^+_c$ &  $q^-_c$ & $A_\Phi$ \\
\hline
 1 & Solenoidal forcing & $yz$ & $t_1$ & 1.74 & 0.70 & 1.77 & 1.69 & 2.032 & ... & ... &  -5 & 0.21 \\
 2 & Solenoidal forcing & $xy$ & $t_1$ & 1.68 & 0.34 & 1.73 & 1.67 & 2.042 & ... & ... &  -7 & 0.21 \\
 3 & Solenoidal forcing & $yz$ & $t_2$ & 1.71 & 0.63 & 1.74 & 1.64 & 2.033 & ... & ... &  -6 & 0.20 \\
 4 & Compressive forcing & $yz$ & $t_1$ & 1.24 & 0.26 & 1.25 & 0.78 & 2.252 & ... & ... &  -6 & 0.49 \\
 5 & Compressive forcing & $xy$ & $t_1$ & 1.35 & 0.31 & 1.38 & 1.10 & 2.213 & 4.52 & ... & ... & 0.56 \\
 6 & Compressive forcing & $yz$ & $t_2$ & 1.31 & 0.17 & 1.33 & 0.98 & 2.254 & 4.12 & ... & ... & 0.57 \\
 7 & Quasi-hydrodynamical & $yz$ & $t_1$ & 1.56 & 0.26 & 1.59 & 1.43 & 2.053 & ... & ... &  -7 & 0.23 \\
 8 & Quasi-hydrodynamical & $xy$ & $t_1$ & 1.61 & 0.35 & 1.65 & 1.56 & 2.075 & ... & ... &  -5 & 0.27 \\
 9 & Quasi-hydrodynamical & $yz$ & $t_2$ & 0.86 & 0.17 & 0.87 & 0.33 & 2.100 & 3.09 & ... & ... & 0.32 \\
10 & High-magnetization & $yz$ & $t_1$ & 1.53 & 0.01 & 1.56 & 1.42 & 2.043 & ... & ... &  -5 & 0.22 \\
11 & High-magnetization & $xy$ & $t_1$ & 1.62 & 0.10 & 1.66 & 1.60 & 2.052 & ... & ... &  -5 & 0.26 \\
12 & High-magnetization & $yz$ & $t_2$ & ... & ... & 1.20 & 0.23 & 2.048 & ... & 14 &  -3 & 0.27 \\
\hline
\end{tabular}
\end{table*}

First, the MFSs of the simulations look quite different, in general, from those of the 
Hi-GAL maps. Particularly on the right side of the curve, for the simulations the largest value 
of $\alpha$, corresponding to the most negative explored $q$, is larger than the one of the 
observations (a more quantitative description of this aspect will be given in 
Section~\ref{quantifying}). Notice also that for some spectra cusps
like those described above, but located at the opposite extreme, appear. In some cases, the cusp 
can not be noticed in the figure because the turnaround point is found close to $q=-10$ and the following
points are very close to those of the ``regular'' part of the MFS. Anyway, in Table~\ref{modmfstab}
we account for the possible occurrence of such an effect for $q$ smaller than a critical $q^-_c$, 
together with possible occurrences of cusps in the $q>0$ portion of the spectrum as well, and with 
other relevant parameters of the MFSs, similarly to Table~\ref{mfstab}. 

Second, remarkable differences are found among the models: the ``compressive forcing'' 
case shows much wider spectra with extended right tails, produced by the presence of large void 
regions already well recognisable in Figure~\ref{modelimage}, second row. 
Their shape and extent are due, in turn, to those original cavities in the 3-dimensional 
synthetic cloud, and to the possible contribution by the foreground medium, depending on the projection 
direction, which is more relevant in presence of a highly inhomogeneous matter distribution.
In this $q<0$ part, in fact, the MFSs of the ``compressive forcing'' maps appear to be more sensitive 
to projection effects than to evolutionary ones, being the $t_2$ curve contained between the 
two $yz$ and $xy$ projections at the time $t_1$. This trend is found also in the
``solenoidal'' case, but within the framework of narrower MFSs, qualitatively more similar to 
observations, at least to the case less influenced by the presence of strong singularities (bright
compact sources), namely $\ell215$. On the contrary, the appearance of bright spots
in the maps of the last two scenarios, ``quasi-hydrodynamical'' and ``high-magnetization'' 
(bottom panels), being the gravitational collapse one of the possible ingredients of these 
models, produces the broadening of the left tail of the MFS at increasing evolutionary 
time ($t_2$); the right tail gets wider as well,
but in this case the evolutionary effect can be confused with the projection effects already
seen for the other scenarios. In short, the MFS of the last two models at time $t_1$
look relatively similar to those of the ``solenoidal forcing'' case, but broaden
for the later epoch $t_2$, characterised by an enhancement of star formation 
activity inside the cloud. Furthermore, since the star formation
efficiency is higher in the ``quasi-hydrodynamical'' case than in the ``high-magnetization'' 
one, the left tail of the MFS at $t_2$ turns out to be much wider in the former than 
in the latter. In Appendix~\ref{simevol} a more systematic analysis of the evolution
of the MFS left tail width with time in presence of gravity in the simulations is provided.

\begin{figure*}
\includegraphics[width=17cm]{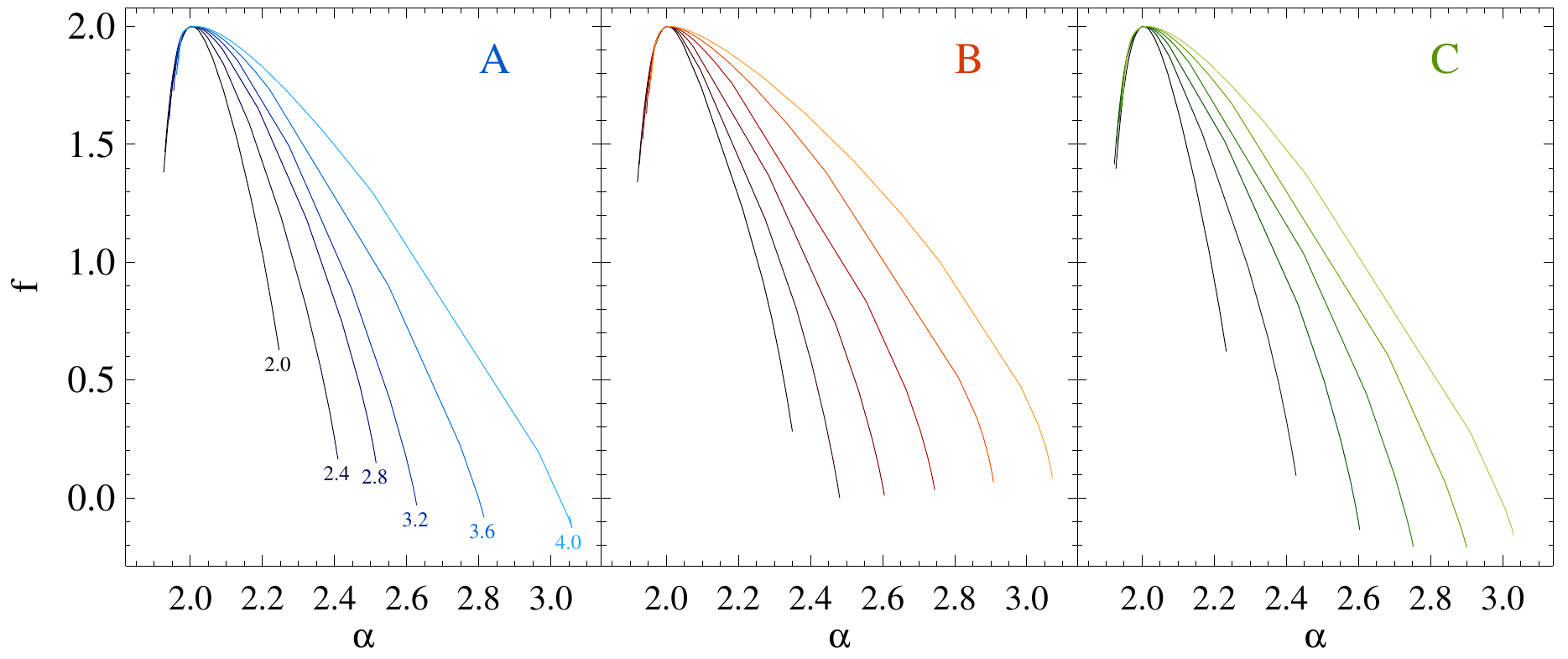} \caption{MFS of the fBm reference images
reported in Figure~\ref{fbmfig}, grouped by image phase distribution (cases ``A'', ``B'',
and ``C'' in left, middle and right panel, respectively). A different color shade, from 
the darkest to the lightest, is used for indicating images with the power spectrum exponent 
$\beta$ ranging from 2 to 4 in steps of 0.4.}\label{allfbmmfs}
\end{figure*}

\begin{table*}
\caption{Main properties of the MFS of the analysed fBm images.}
\label{fbmmfstab}
\begin{tabular}{lccccccccccc}
\hline
Group & $\beta$ & $\alpha_{20}$ & $f_{20}$ & $\alpha_{10}$ & $f_4$ & $\alpha_0$ & $\alpha_{-10}$ &  $q^-_c$ & $A_\Phi$\\
\hline
A & 2.0 & 1.93 & 1.38 & 1.95 & 1.95 & 2.004 & 2.25 & ... & 0.06 \\
 & 2.4 & 1.93 & 1.47 & 1.95 & 1.95 & 2.006 & 2.41 & ... & 0.08 \\
 & 2.8 & 1.94 & 1.61 & 1.96 & 1.95 & 2.009 & 2.52 & ... & ... \\
 & 3.2 & 1.95 & 1.73 & 1.97 & 1.96 & 2.011 & 2.63 & ... & ... \\
 & 3.6 & 1.96 & 1.80 & 1.97 & 1.96 & 2.014 & ... &  -8 & ... \\
 & 4.0 & 1.97 & 1.84 & 1.97 & 1.97 & 2.016 & ... &  -5 & ... \\
B & 2.0 & 1.92 & 1.34 & 1.95 & 1.95 & 2.004 & 2.35 & ... & 0.07 \\
 & 2.4 & 1.93 & 1.42 & 1.95 & 1.95 & 2.005 & 2.48 & ... & 0.08 \\
 & 2.8 & 1.93 & 1.52 & 1.95 & 1.95 & 2.007 & 2.60 & ... & ... \\
 & 3.2 & 1.94 & 1.63 & 1.96 & 1.95 & 2.009 & 2.74 & ... & ... \\
 & 3.6 & 1.95 & 1.72 & 1.96 & 1.95 & 2.011 & 2.91 & ... & ... \\
 & 4.0 & 1.96 & 1.78 & 1.97 & 1.96 & 2.013 & 3.07 & ... & ... \\
C & 2.0 & 1.93 & 1.40 & 1.95 & 1.95 & 2.004 & 2.23 & ... & 0.06 \\
 & 2.4 & 1.92 & 1.42 & 1.95 & 1.94 & 2.006 & 2.43 & ... & 0.08 \\
 & 2.8 & 1.93 & 1.51 & 1.95 & 1.94 & 2.008 & 2.60 & ... & ... \\
 & 3.2 & 1.94 & 1.59 & 1.95 & 1.94 & 2.010 & ... &  -9 & ... \\
 & 3.6 & 1.94 & 1.65 & 1.96 & 1.95 & 2.013 & ... &  -7 & ... \\
 & 4.0 & 1.95 & 1.70 & 1.96 & 1.95 & 2.016 & ... &  -8 & ... \\
\hline
\end{tabular}
\end{table*}

To complete this qualitative description of the computed MFS, let us look at the spectra of
the fBm reference images, with particular emphasis to changes in the MFS as a consequence of
variations of their fractal parameters. In Figure~\ref{allfbmmfs} the MFS, calculated for all the 
18 images generated for this work (see Section~\ref{fbm} and Appendix~\ref{fbmappend}), are shown.
The increase of the power-law exponent $\beta$ clearly produces a systematic increase of the width 
of the $q<0$ part, but also a drift of the extreme point of the $q>0$ part towards larger values of
$f$ and $\alpha$. This right-sided asymmetry indicates, in general, that small scale fluctuations
exhibit a more pronounced multifractal behaviour than large scale ones. Again, let us notice that 
this overall trend does not exclude possible peculiar differences from image to image. Surprisingly, 
even for images generated with the same $\beta$ (then with the same the fractal dimension, according 
to Equation~\ref{dfbm}) we can find differences depending on the choice of the random phases. It is 
possible to ascertain this fact comparing the shape of the MFS of images having the same $\beta$ in 
the three panels of Figure~\ref{allfbmmfs}, but corresponding to different random phase distributions.
Finally, it is to notice that also the MFSs of the fBm images can suffer of the presence of 
cusp-like features, namely at the smallest negative $q$ orders. Similarly to previously considered
classes of images, in Table~\ref{fbmmfstab} the most relevant features of the MFS of fBm sets are
listed.

\subsection{Quantifying the information contained in the MFS}\label{quantifying}

In order to better quantify the information contained in the analysed MFSs, one needs to
establish some meaningful descriptors of the geometry of the spectrum. The literature about 
multifractal analysis abounds with examples of such indicators. In the following discussion,
we use a set of indicators which are independent of each other, and/or are able to highlight
common trends and differences among the analysed sets.

\subsubsection{Dimensional diversity vs maximum singularity strength}
\begin{figure*}
\includegraphics[width=17.5cm]{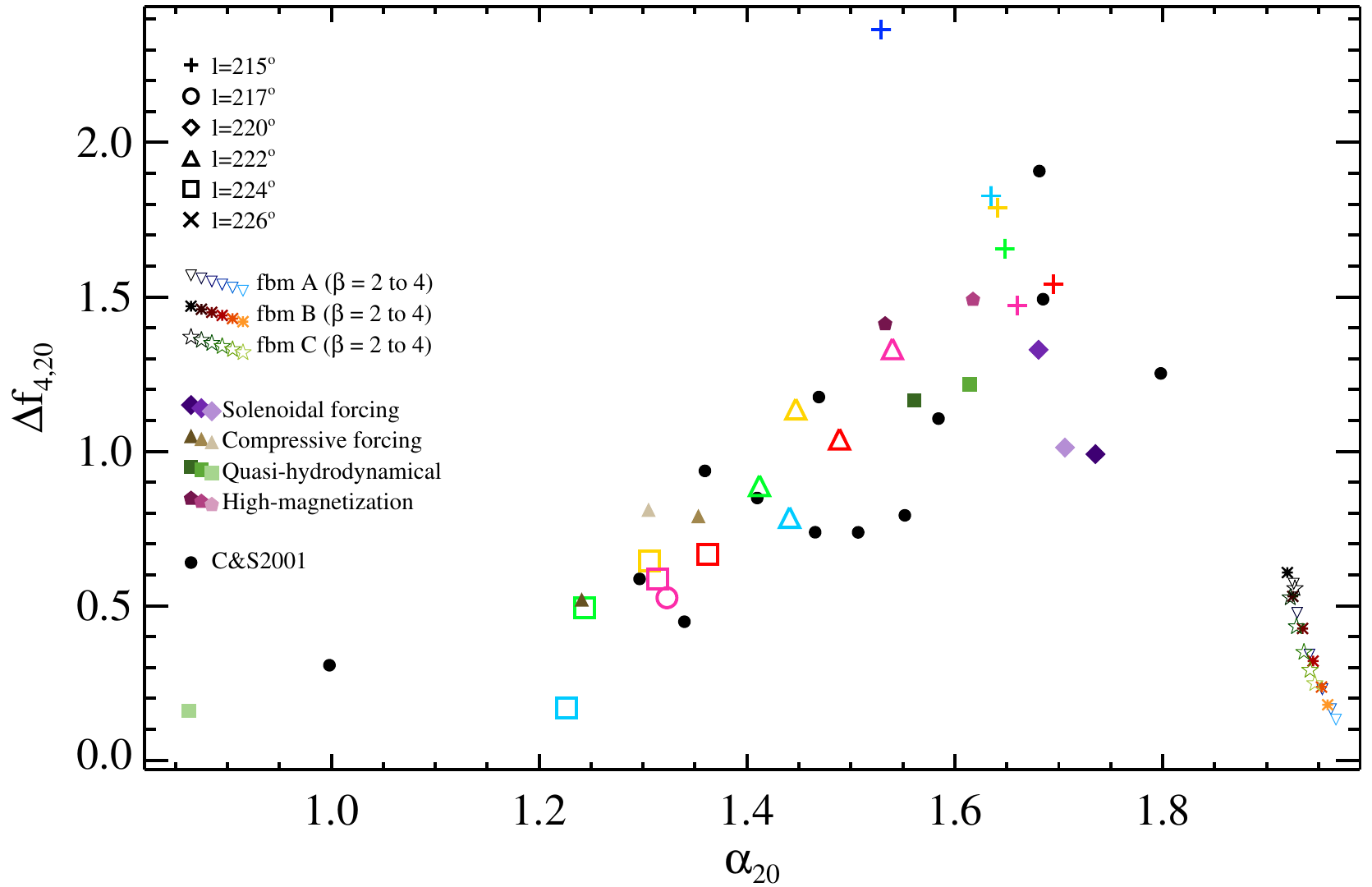} \caption{Plot of ``dimensional diversity'', 
expressed by $\Delta f_{4,20}\equiv f(q=4)-f(q=20)$, versus the smallest computed value of 
the singularity strength $\alpha_{20} \equiv \alpha(q=20)$. Such quantities are available 
only for Hi-GAL maps with $q^+_c\geq 20$ . The correspondence of symbols and colors with
tiles and bands, respectively, is explained in the legend; the symbol for the tile $\ell226$,
although not present in the plot, is introduced since the convention established
here is used also in other following figures. Points representing the the 12 cloud simulations
and the 18 fBm images analysed in this work are plotted as well (symbols are explained in 
the legend). Furthermore, the positions of the 18 fBm reference images (located 
close to the bottom right corner of the diagram) are represented with smaller symbols, using 
color scales (different for the ``A'', ``B'', and ``C'' cases) to identify the different 
explored values of $\beta$. 
Finally, for comparison, points taken from Figure~8 of 
\citet{cha01} and representing multifractal features of their IRAS-based column density 
maps of 13 nearby ($d \leq 160$~pc) star forming regions are reported as filled black
circles (``C\&S2001'', in the legend).}\label{chappellplot}
\end{figure*}

Here we start from the diagnostics adopted by \citet{cha01}, which 
permits a direct comparison with the unique previous case of MFS analysis of 
interstellar clouds (but with the limitation of considering only the left part of the MFS, 
neglecting the negative $q$ orders). These authors used $\alpha_{20}$
as a measurement of the strength of the brightest singularities found in a given map, and 
the $f_4-f_{20}$ offset (hereafter $\Delta f_{4,20}$) as a ``dimensional 
diversity'' to characterise their fields: an image containing isolated, strong and point-like
concentrations will have low values of $f$ for high $q$ orders, while this parameter is 
expected to increase in presence of a large variety of geometries. These authors recognised an
increasing trend in the $\Delta f_{4,20}$ vs $\alpha_{20}$ scatter plot for their sample of 
IRAS-based column density maps of 13 nearby star-forming regions, so that structures with 
strong dominant concentrations (low $\alpha_{20}$) typically have smaller dimensional
diversities (low $\Delta f_{4,20}$). 

We extend this analysis to our data sets (Hi-GAL observations, simulations and fBm 
images) for which it was possible to compute the MFS up to $q=20$ (see Tables~\ref{mfstab} 
and \ref{modmfstab}, respectively). We find in Figure~\ref{chappellplot} that Hi-GAL and 
simulation maps follow the general trend delineated by the maps of \citet{cha01}. 
In particular, among the Hi-GAL tiles, those present at various different wavelengths 
are  $\ell215$, whose different bands occupy the upper part of such trend regardless of 
wavelength (except, again, for the completely peculiar behaviour of the $70~\mu$m map), 
$\ell222$ in the middle, and $\ell224$
at the bottom left end, where the column density map of $\ell217$ is found as well. In practice,
the maps containing the strongest singularities, i.e. the most active star forming sites,
$\ell217$ and $\ell224$, show not only, as expected, lower $\alpha$ values, but also a lower
degree of dimensional diversity. Compared with the observational set of \citet{cha01}, their 
morphology appears, from the statistical point of view, similar to that of their L~134 and 
Cham~1 cases. On the contrary, tiles more quiescent from the point of view of the star ormation, 
such as $\ell222$ and $\ell215$, populate an upper
region of the diagram. The latter, in particular, is found to be close to the positions of 
the fields Oph~N, W, and U of \citet{cha01}, defined by these authors as the most ``space
filling'' in their sample.
The points representing models are spread along the same general trend, as well. Interestingly,
the ``solenoidal'' case is found in the top-right part of the diagram, while the ``compressive'' 
one, characterised by stronger singularities, is found at smaller $\alpha_{20}$. For both of 
them no particular dependence on projection or evolutionary effects is seen. Also, both the 
``quasi-hydrodynamical'' and the ``high-magnetization'' cases, at the time $t_1$ populate the
top-right part of the diagram, but the evolutionary effect is strong in these cases, so that
the corresponding MFS broadening seen in Figure~\ref{modelmfs} is mirrored in this diagram
by a significant decrease of $\alpha_{20}$ at $t_2$, which can be seen for the former model, 
while for the latter it can be guessed but not displayed because $q^+_c=14$.

Finally, a surprising behaviour is seen for the fBm images in this diagram. They occupy a completely 
different region of the plot, corresponding to $\alpha_{20}>1.9$ and $0.1 < \Delta f_{4,20} < 0.6$. 
This means that, compared with both observations and models, these images show at the same time a 
narrower right tail of the MFS, and a low content of fractal diversity. Inside the region occupied 
by these sets, a trend from top-left to bottom-right is observed at increasing $\beta$ (so at 
decreasing fractal dimension). Anyway, all the fBm images appear segregated from the main trend 
of observational maps, and this should impose serious restrictions to the use of the fBm images 
as surrogate of the ISM maps, not yet highlighted in literature.

In fact, the assumed affinity between the fBm images and observations of ISM is based on a 
certain visual similarity and the fact that the power spectrum of the latter ones exhibits a power-law 
behaviour and some randomness in the phase distribution \citep{stu98}. However, such a power-law
is found only over a limited range of scales \citep[e.g.,][]{stu98,sch11}, which is typical for natural 
fractals, and/or different slopes can be found in different ranges of spatial scale 
\citepalias{eli14}. In particular, in \citetalias{eli14} it has been shown how strong departures from a 
single power-law behaviour appear if an original fBm image is manipulated to obtain a more realistic image,
i.e. removing its periodicity, and/or easing off the emission along the borders of the image, and/or 
enhancing the emission in brightest regions to simulate the presence of strong overdensities (their Figure~3).  
Correspondingly, a change in the Fourier phases is expected as well. For example, \citet{bur16}, analysing 
numerical simulations of isothermal compressible turbulence and focusing on information about phases, found 
that degree of coherence in phase distributions (being the coherence equal to 0 in the case of a random 
distribution) depends on some simulation parameters, such as the sonic Mach number. Our present 
empirical analysis, therefore, highlights the need of a further comparative study of the
phases among observational, numerical, and fBm maps.

\subsubsection{Peak position}

\begin{figure}
\includegraphics[width=8.5cm]{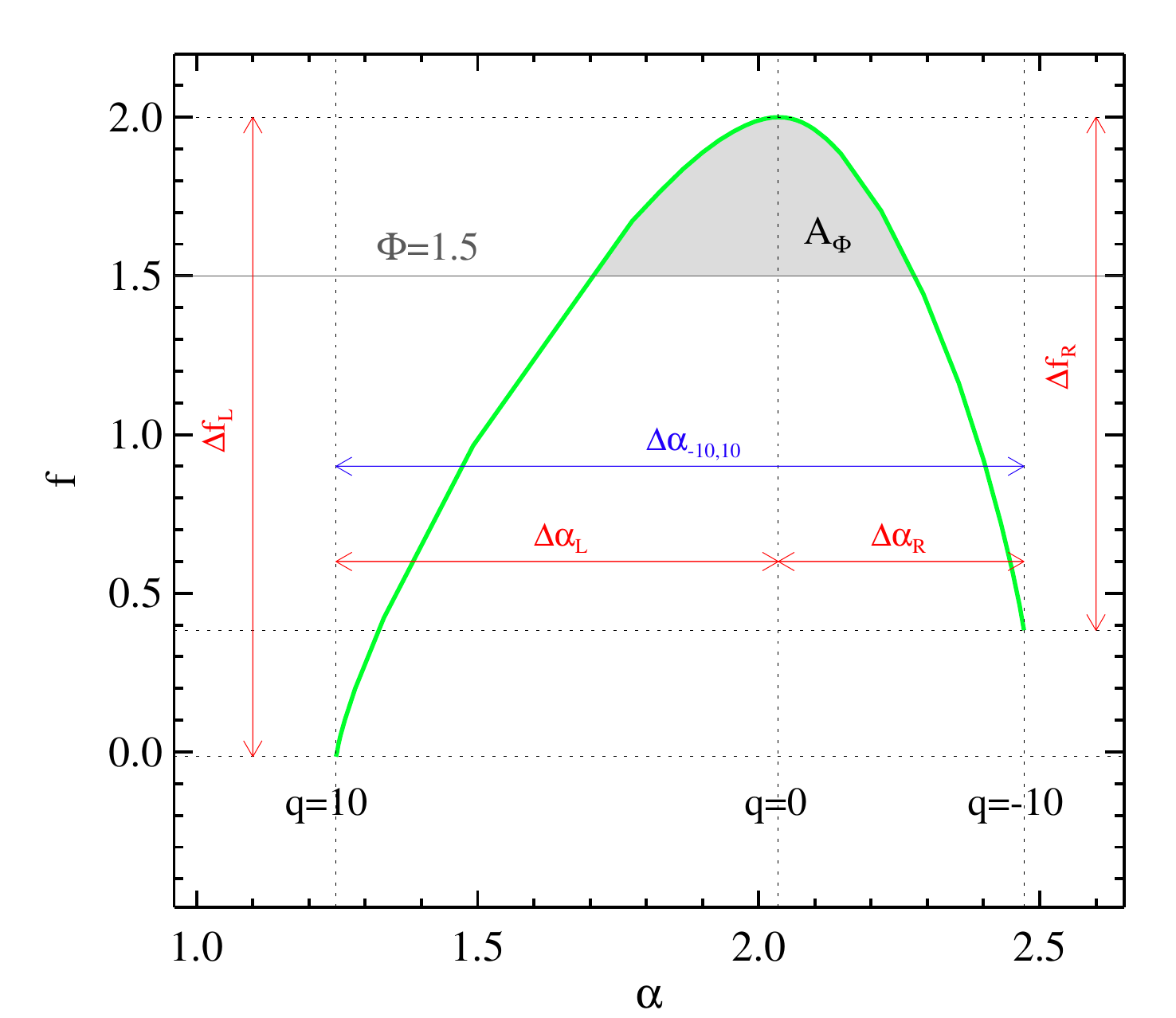} \caption{Scheme of descriptors used to 
quantitatively characterize the MFS in this work, and introduced in the text. The example
MFS (green curve) is the one of the $\ell224$ map at 250~$\mu$m, also contained in 
Figure~\ref{mfsspectra}.}\label{quantmfs}
\end{figure}

The analysis contained in the following represents an expansion of the
\citet{cha01} approach, thanks to the introduction of further diagnostics to 
describe the shape of the MFS. A graphic illustration of these quantities is also given, 
where possible, in Figure~\ref{quantmfs}.

Let us start with $\alpha_0$, namely the abscissa of the MFS peak: it is lower if the signal 
is uncorrelated and the underlying process ``loses fine structure'', i.e. the dominant 
fractal structure has more signal at larger fluctuations, since fine fluctuations become less
frequent, so the object becomes more regular in appearance. In Table~\ref{mfstab} a systematic 
increase\footnote{The differences among $\alpha_0$ values are generally found on the second or 
the third decimal digit (for this reason the precision of the 
$\alpha_0$ values quoted in Tables~\ref{mfstab}, \ref{modmfstab}, and \ref{fbmmfstab} is increased with
respect to other analogous parameters). Despite this, they show interesting systematic trends.} is seen for 
$\alpha_0$ from 70 to 500~$\mu$m for any Hi-GAL tile. In particular, the largest gaps are seen between 70
and 160~$\mu$m, and from 160~$\mu$m to SPIRE wavelengths, while the column density maps show  
$\alpha_0$ values close to those of the SPIRE maps, with no correspondence to a particular wavelength.
The observed behaviour can be ascribed to the increasing degree of ``structure'' at increasing 
wavelength, namely a gradual enhancement of diffuse emission compared to isolated strong singularities.
Furthermore, the most quiescent single tiles ($\ell215, \ell220$) show lower $\alpha_0$ compared 
to the others; in particular, at SPIRE
wavelengths, they are characterized by $\alpha_0< 2.015$, while $\alpha_0> 2.020$ for the remaining tiles.   
No systematic trends are found, instead, among different simulations (Table~\ref{modmfstab}), except 
the remarkably large $\alpha_0$ values for the ``compressive forcing'' case, corresponding 
to a significantly more correlated signal, as already found, with respect to the ``solenoidal forcing'' 
case, by \citet{fed09} through the structure function analysis.
Finally, a little but systematic increase of $\alpha_0$ at increasing $\beta$ is seen for the fBm images 
(Table~\ref{fbmmfstab}); notice indeed that for this class of objects, starting
from the case $\beta=0$ (white noise), corresponding to a totally uncorrelated signal, correlation 
increases with increasing $\beta$. 

\subsubsection{MFS width}
\begin{figure}
\includegraphics[width=8.5cm]{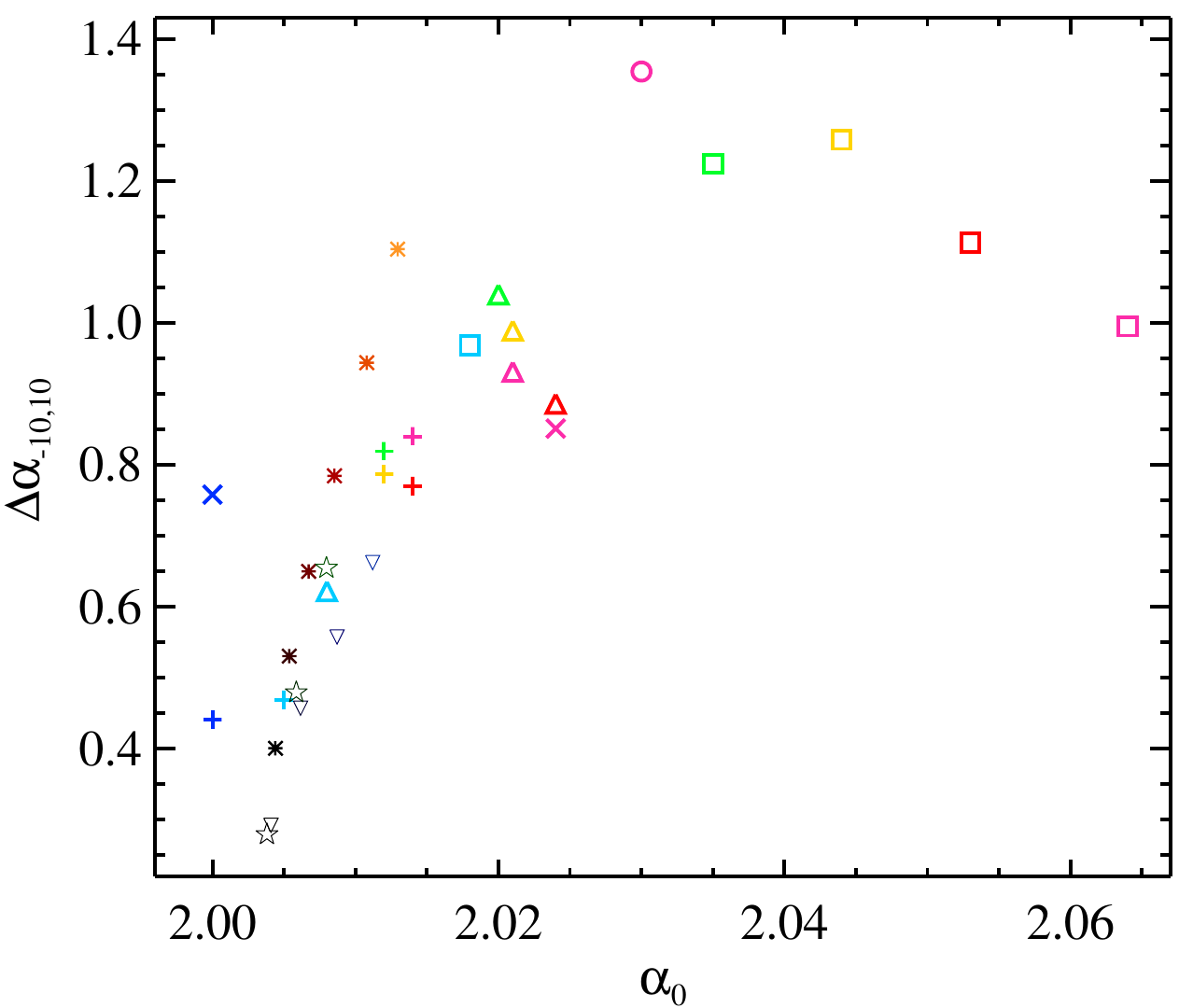} \caption{Plot of the MFS ``amplitude'' 
$\Delta\alpha_{-10,10}$ versus the peak position $\alpha_0$ for the maps studied in this
work for which a reliable $\alpha$ values in the the $10\leq q \leq 10$ range was derived. Data 
corresponding to the three turbulent ISM simulations with available $\Delta\alpha_{-10,10}$
(see Table~\ref{modmfstab}) are all located at $\alpha_0>2.1$ and $\Delta\alpha_{-10,10}>2.2$ 
and are not displayed here not to overly compress the plot. The 
symbols are the same introduced in the legend of Figure~\ref{chappellplot}.}\label{apeak}
\end{figure}

A second relevant quantity is the MFS width, which expresses the degree of multifractality 
of the investigated set. The wider the range, the more multifractal are the fluctuations in 
the image. We express it here through the $\Delta\alpha_{-10,10}\equiv \alpha_{-10}-\alpha_{10}$ 
parameter \citep[see, e.g.,][]{mac07}. In Figure~\ref{apeak}, this quantity is plotted versus
$\alpha_{0}$ for those maps having a MFS not involving cusps in $-10\leq q \leq 10$. The 
considerations written above about $\alpha_{0}$ are easily recognizable in the abscissae of
the points in the plot but, in addition, an interesting trend is seen between the two
plotted quantities, with the degree of multifractality $\Delta\alpha_{-10,10}$ generally increasing 
at increasing degree of complexity expressed by $\alpha_{0}$. Such a trend is more easily 
recognizable for fBm images at increasing $\beta$, although confined to a shorter range of
$\alpha_{0}$, than for Hi-GAL maps, whose corresponding points are more scattered.

\begin{figure}
\includegraphics[width=8.5cm]{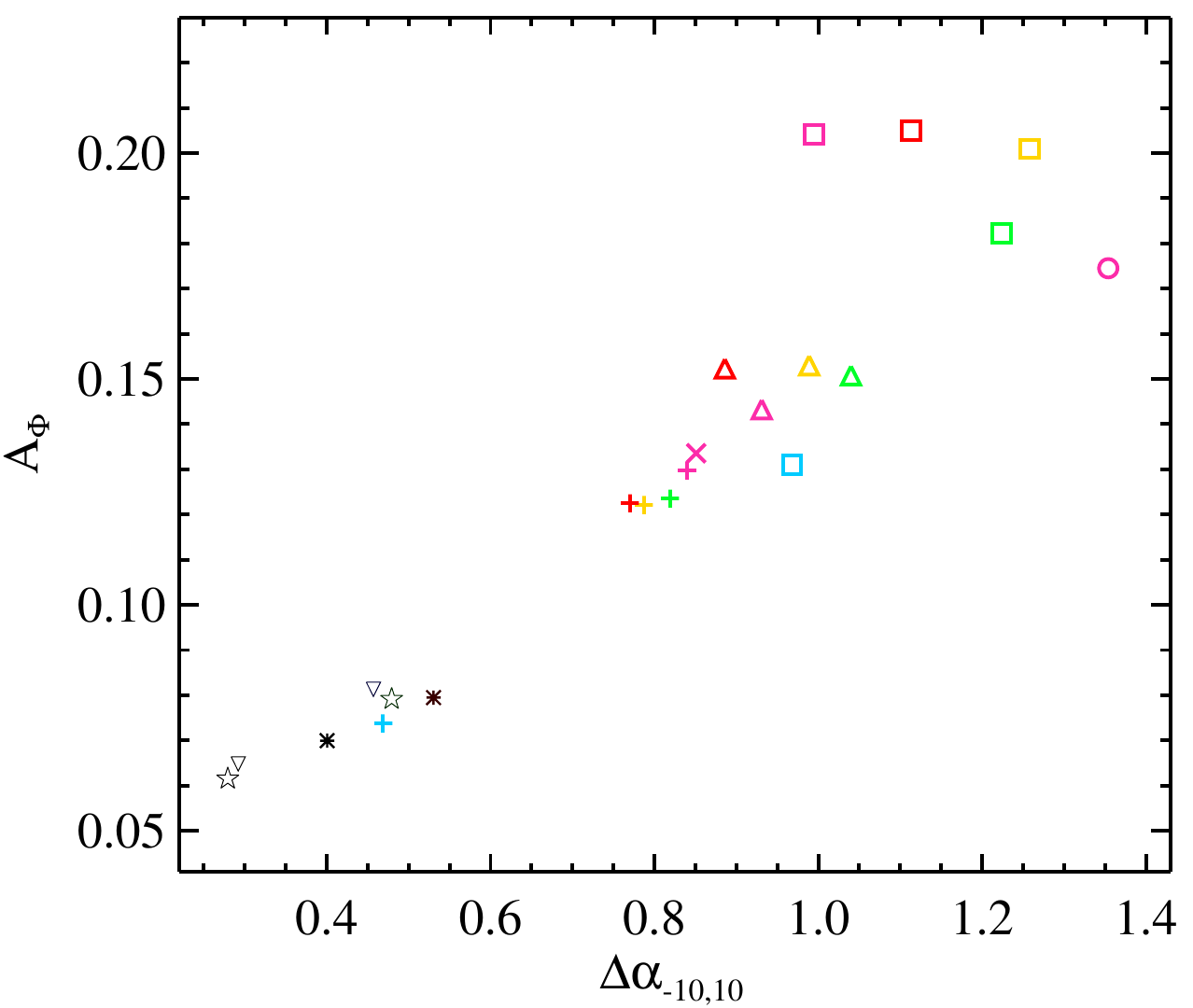} \caption{Plot ot the area $A_{\Phi}$ under the 
MFS and above $f=1.5$ versus $\Delta\alpha_{-10,10}$ for maps for which both quantities can be 
derived. The symbols are the same introduced in the legend of 
Figure~\ref{chappellplot}.}\label{aphi}
\end{figure}

Another way to measure the amplitude of a MFS is to compute the area $A_\Phi$ 
delimited by the MFS curve and a given horizontal cut at a level $f=\Phi$:
\begin{equation}
A_\Phi=\int_{\alpha_{\Phi,1}}^{\alpha_{\Phi,2}} [f(\alpha)-\Phi]\,d\alpha\;,
\end{equation}
where $\alpha_{\Phi,1}$ and $\alpha_{\Phi,2}$ represent the abscissae of the
two points of the MFS for which $f=\Phi$.
The area $A_\Phi$, illustrated in Figure~\ref{quantmfs} as a grey region, can be evaluated 
only for MFSs whose left and right tails are both intersected 
by the cut at $f=\Phi$, which is formally expressed by the condition 
$\alpha_{\textrm{min}}<\alpha_{\Phi,1}<\alpha_{\Phi,2}<\alpha_{\textrm{max}}$. This also accounts for 
the presence of possible cusp-like behaviour in the spectrum. The $A_\Phi$ obtained (choosing $\Phi=1.5$)
for Hi-GAL fields, fBm sets and simulations are quoted in Tables~\ref{mfstab}, \ref{modmfstab}, and \ref{fbmmfstab},
respectively.

The correlation between $A_\Phi$ and $\Delta\alpha_{-10,10}$ is investigated 
in Figure~\ref{aphi}, displaying a roughly linear correlation between these two 
descriptors, as expected, since an increase of the MFS width should correspond to an increase of $A_\Phi$, 
being the ordinate of the MFS peak constant.
Larger departures from this behaviour are seen at large values of these quantities, especially 
for 350~$\mu$m, 500~$\mu$m and column density images of $\ell224$. This indicates a particularly 
``swollen'' shape of MFS, quite well recognisable in the $\ell224$ panel of Figure~\ref{mfsspectra}, 
especially for the colum density map. As in the case of Figure~\ref{apeak}, in Figure~\ref{aphi} the points 
corresponding to simulations (three available cases: \#5, \#6, and \#9 of Table~\ref{modmfstab}) 
are not shown, being all of them located at $\Delta\alpha_{-10,10}>2.2$ and $A_\Phi>0.3$, that is to 
say far from the region of the plot populated by observational data sets, highlighting a different
structure compared with real data. In particular, since both $\Delta\alpha_{-10,10}$ and $A_\Phi$
are descriptors of the degree of multifractality in the signal, this indicates a generally wider variety 
of structures than observed in Hi-GAL tiles.

\begin{figure}
\includegraphics[width=8.5cm]{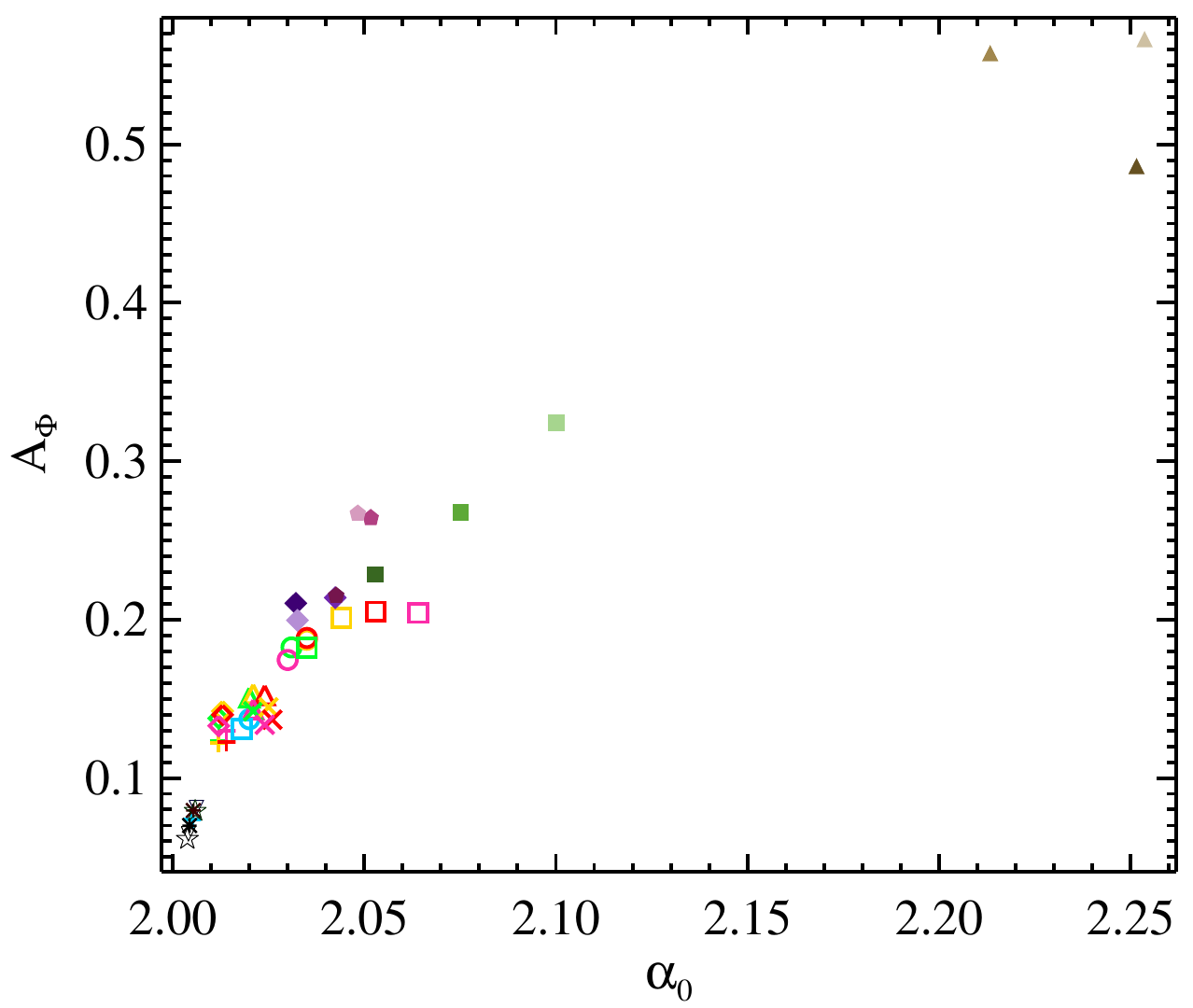} \caption{Plot ot the area $A_{\Phi}$ under the MFS 
and above $f=1.5$ versus the peak position $\alpha_0$ for maps for which $A_{\Phi}$ can be derived. 
The symbols are the same introduced in the legend of Figure~\ref{chappellplot}.}\label{aphivspeak}
\end{figure}

In our case, the benefit of using $A_\Phi$ instead of $\Delta\alpha_{-10,10}$ to represent the MFS 
amplitudes consists of a larger number of useful maps. In Figure~\ref{aphivspeak} the $A_\Phi$ versus 
$\alpha_0$ plot confirms, based on a larger sample of cases, the trends seen in Figure~\ref{apeak}.
The plot appears less scattered than that of Figure~\ref{apeak} with regard to positions
of Hi-GAL and fBm images, while the positions of the simulations cover a wide range towards larger values. 
The ``solenoidal'' and the $t_1$, $yz$ maps of both the ``quasi-hydrodynamical'' 
and ``high magnetization'' scenarios are the closest to the observational points, mostly to the ones 
corresponding to SPIRE maps of $\ell224$ and $\ell217$, that is to say that, from the point
of view of the MFS amplitude, they show a degree of multifractality similar to that of the most
actively star forming regions in our sample.

\subsubsection{Left vs right tail width}\label{mfswidths}
\begin{figure}
\includegraphics[width=8.5cm]{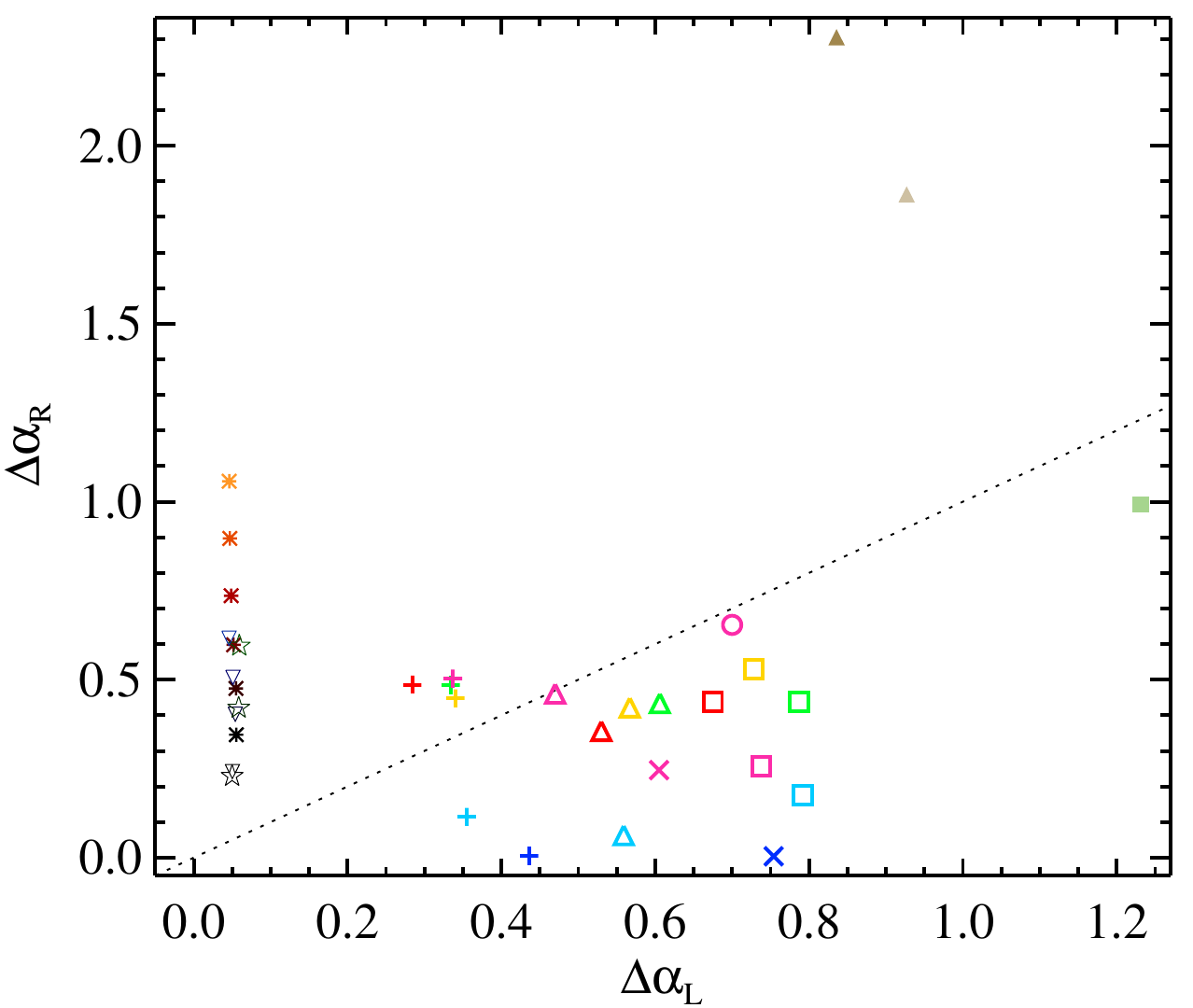} \caption{Plot ot the MFS left versus right tail amplitude 
for maps for which it is possible to derive both quantities. The symbols are 
the same introduced in the legend of Figure~\ref{chappellplot}. The dotted line represents
the bisector, to facilitate the distinction between left-skewed (below the line) and right-skewed
(above the line) MFSs, respectively.}\label{a10a10}
\end{figure}

The diagnostics used in Figures~\ref{apeak}, \ref{aphi}, and~\ref{aphivspeak} consider the MFS
as a whole, neglecting information contained in possible asymmetries. 
In fact, the left tail of the MFS represents the heterogeneity (broad tail) or uniformity
(narrow tail) of the high values distribution, as well as the right tail is related to 
heterogeneity/uniformity of the low values distribution
 \citep[e.g.,][]{pav13,def17}. Strong asymmetries are observed in most of the spectra analysed
 in this paper, with the extreme case represented by PACS 70~$\mu$m maps, discussed above. 
 
In order to make use of the MFS 
asymmetry, single values of $\alpha$ on the two sides of the MFS (for example, $\alpha_{10}$ and 
$\alpha_{-10}$, to follow the same approach as above) should be considered \citep{def17}. In 
Figure~\ref{a10a10} the two widths
$\Delta \alpha_L\equiv \alpha_0-\alpha_{10}$ and $\Delta \alpha_R\equiv \alpha_{-10}-\alpha_0$, 
are plotted. First, a strong segregation in abscissa is visible between the group of fBm sets
(which are right-skewed and for which the width of the left MFS tail $\alpha_{10}$ remains nearly 
constant, see also Table~\ref{modmfstab}) and that of Hi-GAL maps. The latter is internally more 
composite: the most quiescent, $\ell215$, is found on the left while 
the most active, $\ell224$, on the right, as a consequence of the leftward broadening of the MFS in presence
of bright compact sources (cf. Figure~\ref{mfsspectra}). Along the vertical coordinate, a systematic
increase is seen from short to long \textit{Herschel} wavelengths, while the column density maps appear
mixed with SPIRE maps but with no specific trend. 

For the fBm imags, the increase of $\Delta \alpha_R$ at increasing $\beta$ corresponds to 
the progressive rightward broadening of the MFS seen in Figure~\ref{allfbmmfs}. 

About the simulations, as in Figure~\ref{aphi} only
three cases (with a MFS very different in general from that of observations) survive the constraint
of absence of possible cusps, and populate a very different plot region, with 
both coordinates larger than those of observations. The lack of other nine cases
prevents us from completing here quantitatively the qualitative description
of Figure~\ref{modelmfs}. We postpone the analysis of implications of the physics contained in the 
models to Section~\ref{dqanalysis}, in which an equivalent approach, but in terms of generalised fractal 
dimension, is discussed.

\subsubsection{Vertical vs horizontal balance}

\begin{figure}
\includegraphics[width=8.5cm]{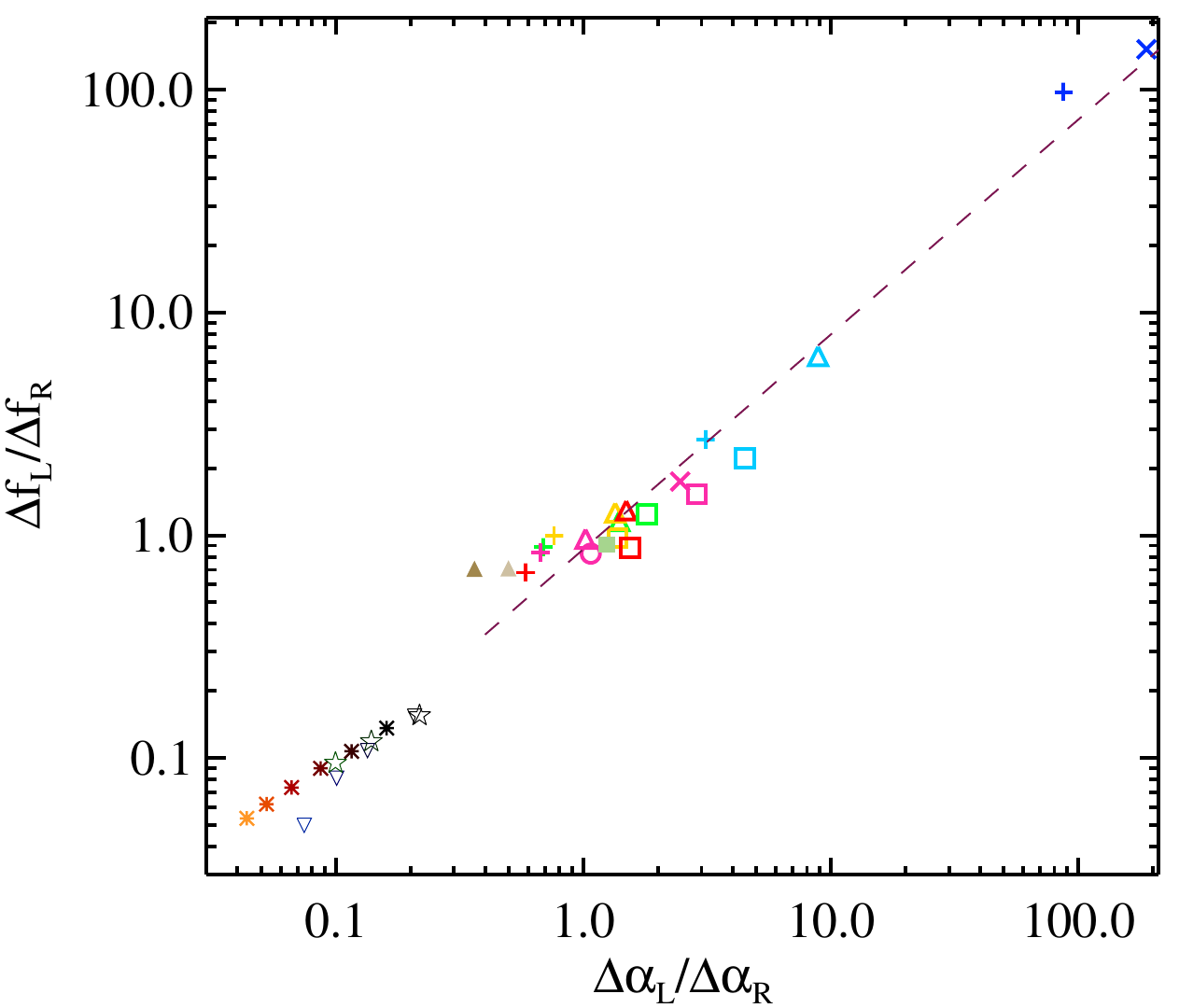} 
\caption{Plot ot the MFS vertical versus 
horizontal balance (estimated involving orders $q=-10,10$) for maps for which it is possible to 
derive both quantities. The symbols are the same introduced in the legend of Figure~\ref{chappellplot}.
The grey dashed line represents the power-law best fit to the locii of the Hi-GAL maps.}
\label{asymmetry}
\end{figure}

The degree of asymmetry of a MFS can be expressed by the ratio between the left and right tail
widths $\Delta \alpha_L/\Delta \alpha_R$ \citep[e.g.,][]{dro15,def17}, with a left-skewed 
spectrum corresponding to the dominance of extreme events, and a right-skewed one denoting 
relatively strongly weighted high fractal exponents, which correspond to fine structures
\citep[e.g.,][]{krz17}. In order to capture the general proportions of the MFS, this horizontal balance 
can be analysed jointly with the vertical one based on the extents along the $f$ direction, 
namely $\Delta f_L/\Delta f_R\equiv (f_0-f_{10})/(f_0-f_{-10})$. For example, an increase of 
the ``fractal divergence'' numerator $\Delta f_L$ is shown by destruction of uniformity into 
a wide range of fractal dimensionality. The vertical versus horizontal balance plot for the 
available MFSs is shown in Figure~\ref{asymmetry}, in which a direct correlation between the two 
is seen, in general. This implies that an asymmetry in the MFS is typically composed
by the predominance of a tail of the spectrum on the other one both in width and in total 
vertical extent. However, in this general trend different sub-regions can be easily recognised:
in the left-bottom part of the diagram ($\Delta \alpha_L/\Delta \alpha_R<0.2,
\Delta f_L/\Delta f_R<0.2$) the fBm images are located, as a consequence of their 
strongly right-skewed MFS. On the contrary, the SPIRE and column density Hi-GAL fields are
found to be mildly right-skewed ($\ell215$) or left-skewed ($\ell217, \ell222, \ell224$). 
The PACS $160~\mu$m maps populate the right-top of this group, while the $70~\mu$m maps
are located very far, at the top right end of the diagram, due to the disproportion between 
the left and the right branch of the MFS in these cases. The trend followed by the Hi-GAL
maps can be fitted to a power-law relation, with 
$\Delta f_L/\Delta f_R=(0.87\pm0.07)\times(\Delta \alpha_L/\Delta \alpha_R)^{(0.96\pm0.04)}$,
i.e. an almost linear behaviour. 
Finally, two out of the three displayed models show a departure from the typical behaviour 
of the observational ISM maps, being located at lower values of $\Delta \alpha_L/\Delta \alpha_R$.

One of the main goals of this work is to show how the extension of the multifractal analysis
of the ISM to negative $q$ orders permits a more refined classification of the images through
their MFS. The tools used to produce Figures~\ref{apeak} to~\ref{asymmetry} confirm 
that the ISM maps show common trends as the one revealed in Figure~\ref{chappellplot} only based
on the positive $q$ portion of the MFS. However they reveal further relations (e.g., 
the behaviour of the peak position, the peculiarities of the PACS maps, the differences between 
observational and simulated maps, etc.) as well, which represent new useful tools we suggest for a 
thorough statistical analysis of ISM maps.

%\subsection{Proposing a metrics for comparison with models}\label{metrics}
%The use of various geometrical features of the MFS considered above can be resumed in a quantitative 
%overall metrics which allows a direct comparison between observational and simulated column
%density maps. First of all, to include into this comparison a larger number of both Hi-GAL tiles and 
%model types, we decide here to quantify the elongations of the MFS through points estimated
%at $q=-5$ and 5 rather than at -10 and 10, respectively, to attenuate the impact of cusp-like
%features in the MFS on the global statistics (in Appendix XXXXXXXX we discuss the suitability of this
%choice). Therefore we base our metrics on a set of five parameters: $w=[\alpha_0, \Delta\alpha_{-5,0}, 
%\Delta\alpha_{0,5}, \Delta f_{-5,0}, \Delta f_{0,5}]$. Using the suffix ``HG'' for the Hi-GAL
%images and ``mod'' for the simulations, we define the generic ``distance'' $\gamma$ between 
%a generic map of the former type and of the latter type as:
%\begin{equation}
%\gamma=\sqrt{\sum_{i=1}^5\left[\frac{w_{\mathrm{HG},i}-w_{\mathrm{mod},i}}{w_{\mathrm{mod},i}}\right]^2}
%\end{equation}
%Such distance can be used to identify possible cases of similar MFSs.  
%XXXXXXXXXXXX

\subsection{Generalised fractal dimension approach}\label{dqanalysis}
In this section, the alternative view provided by the generalised fractal dimension approach is presented.
On the one hand, since this representation is practically equivalent to the MFS one through 
Equations~\ref{legendrea} and~\ref{legendref}, we will not comment in detail the behaviour of $D_q$, but 
mostly and briefly highlight how features found with this approach are already recognisable in the MFS. 
On the other hand, this approach permits a more direct and intuitive comparison with the mono-fractal 
properties of the investigated sets.

\begin{figure*}
\includegraphics[width=17cm]{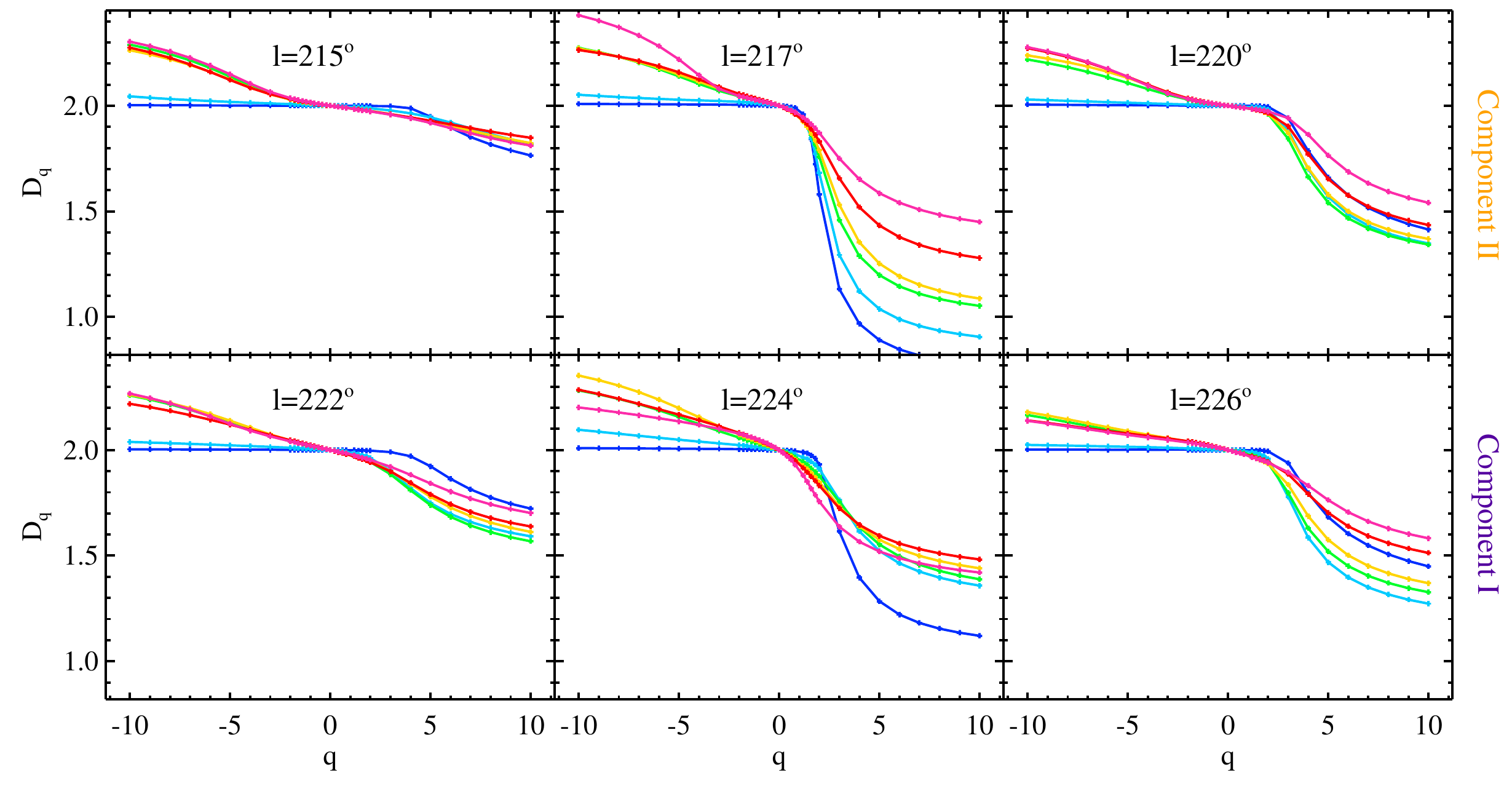} \caption{Generalised fractal dimension
$D_q$ versus order $q$ for the Hi-GAL maps analysed 
in this work, ordered by tile and color-coded as in Figure~\ref{mfsspectra}.}\label{mfsdq}
\end{figure*}

In Figure~\ref{mfsdq} the $D_q$ curves are shown for each Hi-GAL tile and each band (plus the column 
density maps), to facilitate the recognition of possible common trends. The most evident features are:
\begin{itemize}
\item A clear asymmetry is found for all curves between the negative and the positive $q$ regions
(concave upward in the former, and downward in the latter). Its extent mirrors the degree of asymmetry 
typically found for these maps between the right and the left part of the MFS.
\item The maps at 70~$\mu$m show in all cases a peculiar behaviour: for $q<0$, $D_q$ is practically 
constant around $\approx 2$, which corresponds to an uniform plus random noise distribution. This
is seen also in the MFS analysis as an extremely short right tail of the MFS (cf. Figure~\ref{mfsspectra}). 
In contrast, these maps can exhibit very bright concentrations of emission corresponding to star 
forming objects/regions, which are enhanced by the $q>0$ analysis. In fact,
mainly for the tiles with the highest content of star formation, namely $\ell217$ and $\ell224$, the scaling 
of $D_q$ is far from being constant and, instead, the 70~$\mu$m band shows the largest departures
from the typical value of 2, reaching $D_q < 1$ in the case of $\ell217$. This corresponds to the 
enlarged left tails of the MFS observed for these two tiles at this band.
%More in general, the maps at 70~$\mu$m show
%a very different behaviour with respect with those at other bands: as already highlighted in
%\citetalias{eli14}, due to their morphological properties described above they generally depart 
%from a fractal behaviour, as it can be confirmed also by the present multifractal analysis. They
%are considered anyway in the following of the paper, but principally to highlight their differences
%from the maps at other bands, which instead exhibit a structure with a more pronounced fractal
%behaviour.
\item As seen also for the MFS, the maps at 160~$\mu$m show a behaviour which is intermediate between
that of the 70~$\mu$m and that of the SPIRE (and column density) ones, which, in turn, show a more
similar behaviour among them.
\item The tiles $\ell217$ and $\ell224$ show the largest differences, not only at 70~$\mu$m (as 
pointed out above) between fractal dimensions of negative and positive orders, and this indicates
a remarkable departure from a simple fractal behaviour.
\end{itemize}

\begin{figure}
\includegraphics[width=8.5cm]{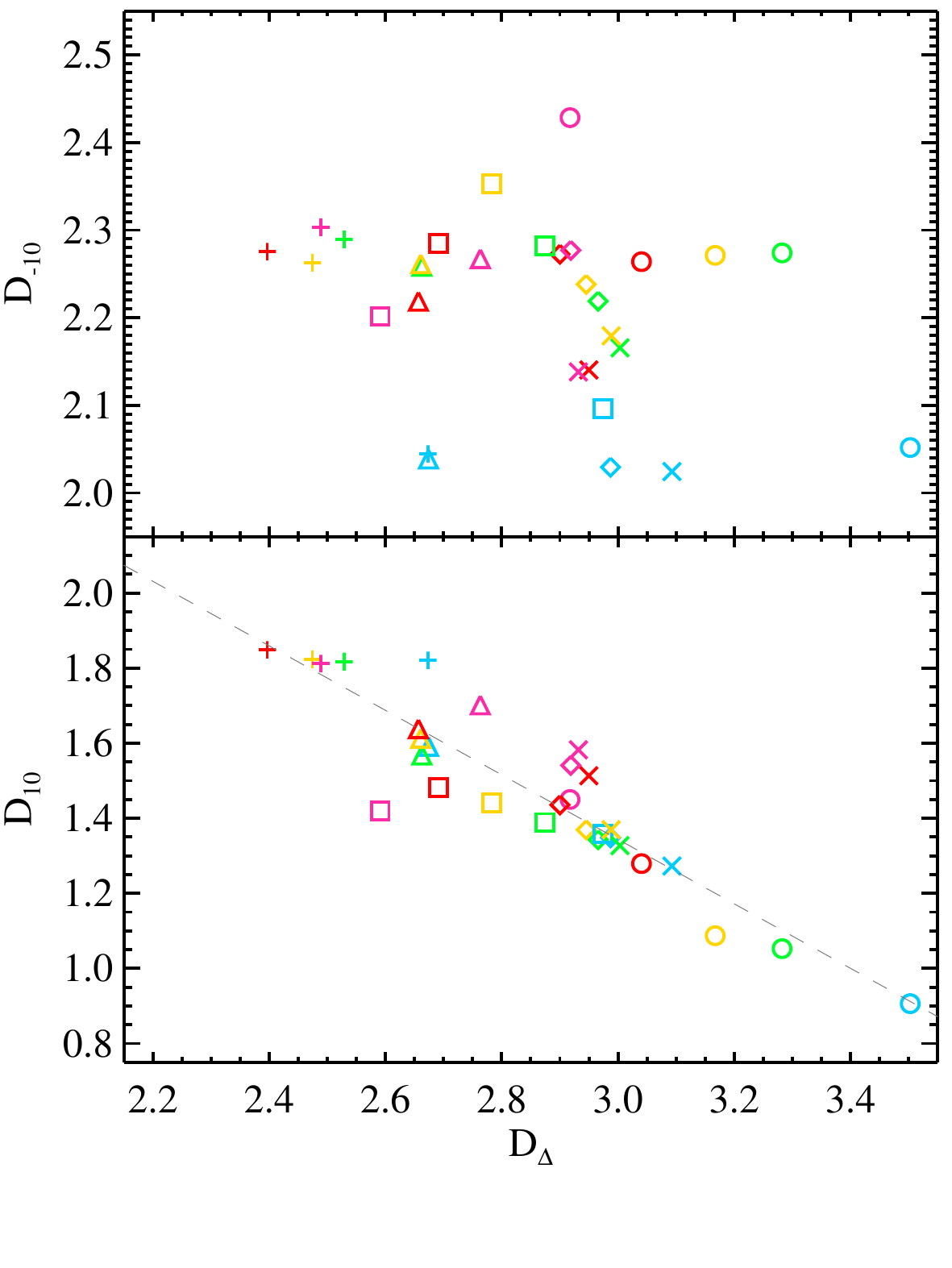} \caption{Plot of the fractal dimensions of order $q=-10$ 
(top panel) and 10 (bottom panel) versus the mono-fractal dimension $D_\Delta$ estimated through
the $\Delta$-variance technique (see Appendix~\ref{dvarapp}) for the six Hi-GAL fields studied in
this paper. Symbols and colors are the same coded in Figure~\ref{chappellplot}, with the addition 
of the $\ell226$ maps, indicated with an asterisk. The 70~$\mu$m maps are not shown, being $D_\Delta$ 
unrealiable for these maps \citepalias[see][]{eli14}.}\label{fracdims}
\end{figure}

Concerning the last point, it is interesting to explore how a mono-fractal indicator, such as the 
fractal dimension estimated as in \citetalias{eli14}, is correlated with the generalised fractal dimensions
estimated here. To do this, the fractal dimensions have been derived for the $\ell215$ and 
$\ell226$ Hi-GAL maps, not included in \citetalias{eli14}, as well as for the
remaining tiles because of their different size (see Section~\ref{higal}). 
In Appendix~\ref{dvarapp} the method used to calculate the fractal dimension $D_\Delta$ for all tiles
at all wavelengths through the $\Delta$-variance technique is described, and the obtained
values are reported (Table~\ref{dvartab}). We use these values to build Figure~\ref{fracdims}, in which
$D_{-10}$ and $D_{10}$ are plotted versus $D_\Delta$ (top and bottom panel, respectively), as
representative cases of negative and positive $q$ dimensions, respectively. First, some considerations
are in order about the distribution of $D_\Delta$: in \citetalias{eli14} it was found that, inside
a given component, tiles rich in star forming regions ($\ell224$ for component~I and 
$\ell217$ for component~II, respectively) generally show, at each band, a larger $D_\Delta$ than 
those containing more quiescent ISM (with an unique exception given by the fractal dimension of the
column density map of $\ell222$, larger than that of $\ell224$). Here this behaviour is 
confirmed\footnote{See Appendix~\ref{dvarapp} for the discussion of the consistency between $D_\Delta$
values obtained in this work and in \citetalias{eli14} for the tiles in common.} by the high 
$D_\Delta$ found for the active field $\ell226$ (component~I) and, on the opposite side, for the 
low ones found for the low-emission field $\ell215$ (component~II). This trend is mixed with 
another one consisting in a systematic decrease (from very evident for $\ell224$ to practically
absent for $\ell222$) of $D_\Delta$ with wavelength from 160 to 500~$\mu$m. 

That being said, the plot of $D_{-10}$ versus $D_\Delta$ shows no correlation between these two dimensions,
being the former relatively well confined: $D_{-10} \lesssim 2.1$ for maps at 160~$\mu$m and 
$2.1 \lesssim D_{-10} \lesssim 2.45$ for SPIRE and column density maps. Instead, the $D_{10}$
dimension shows a linearly decreasing trend at increasing $D_\Delta$, which can be represented 
by a linear fit: $D_{10}=(3.92\pm 0.21)-(0.86\pm 0.07)\times D_\Delta$. In practice, this highlights that,
at least for the investigated Hi-GAL fields, the mono-fractal dimension is mostly sensitive to the 
scaling of large fluctuations and strong singularities.

\begin{figure}
\includegraphics[width=8.5cm]{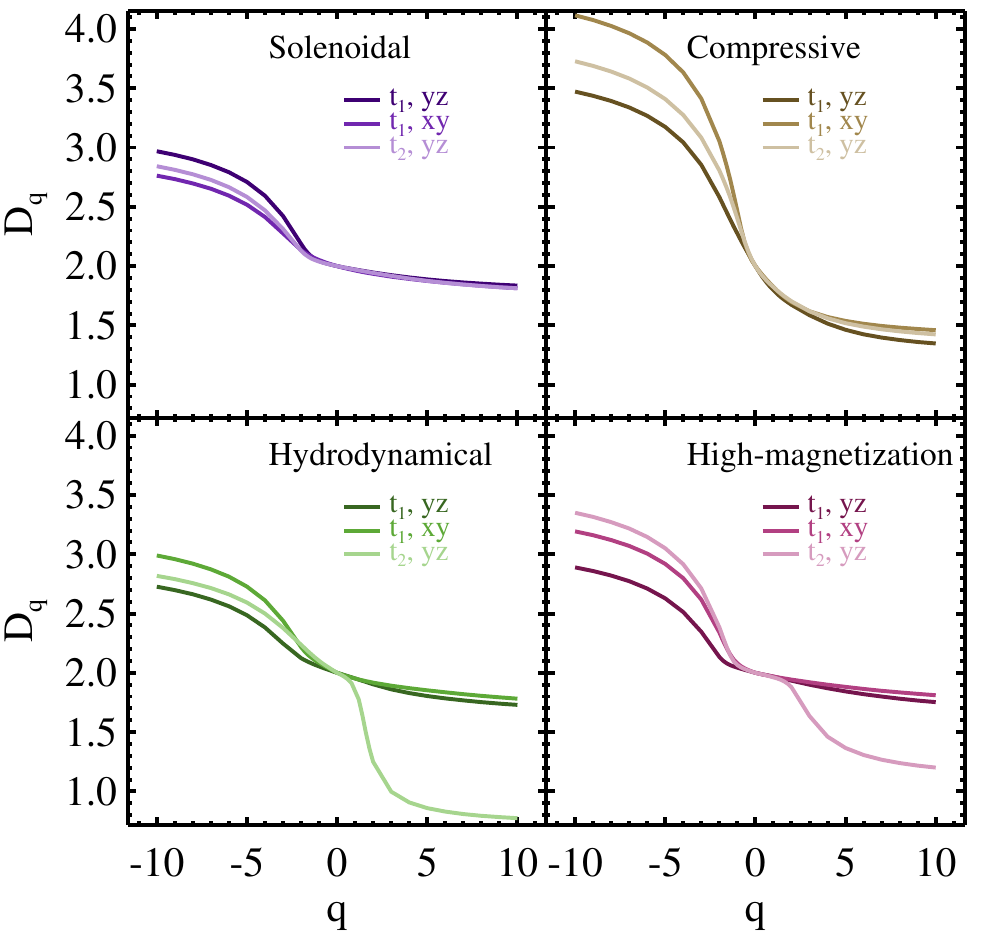} \caption{Generalised fractal dimension $D_q$ versus order 
$q$ for the column density maps produced by the numerical simulations analysed 
in this work, arranged by model and color-coded as in Figure~\ref{modelmfs}.}\label{dqmodel}
\end{figure}

\begin{figure*}
\includegraphics[width=17cm]{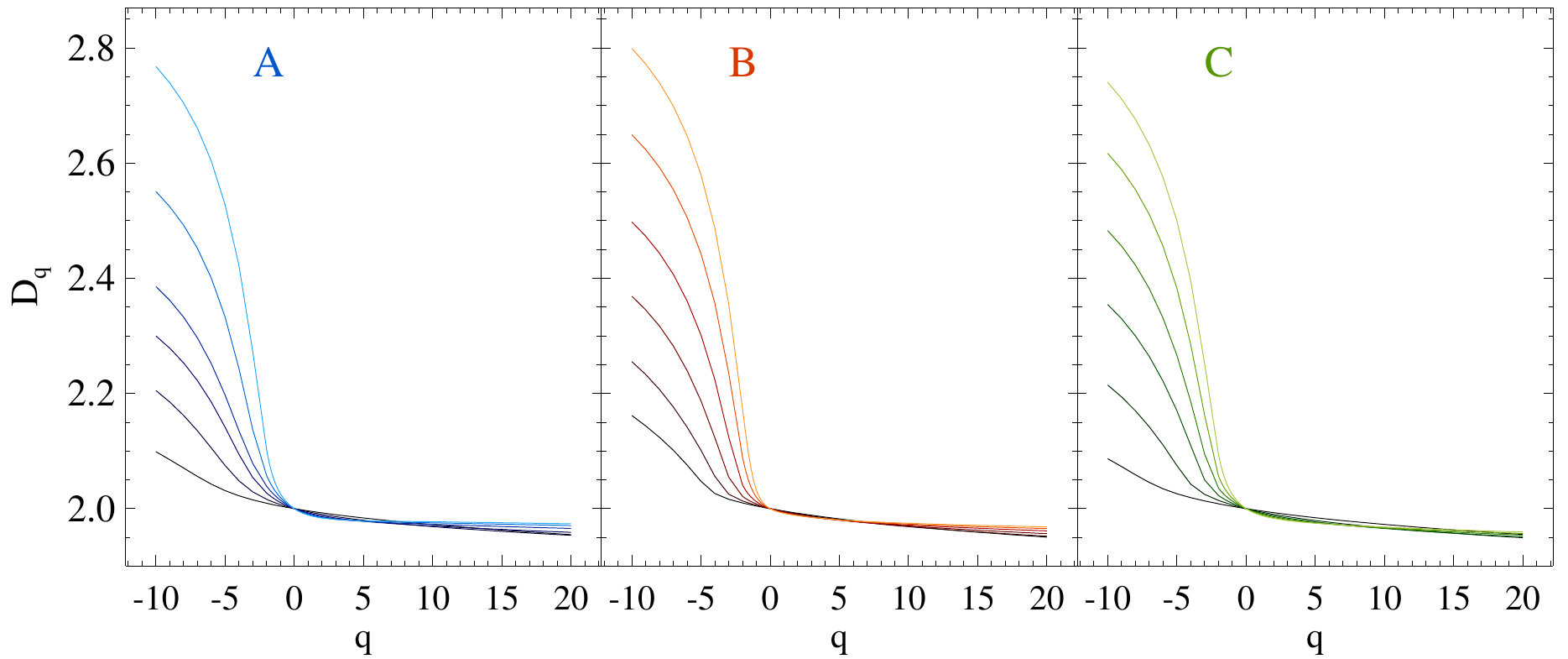} \caption{Generalised fractal dimension
$D_q$ versus order $q$ for the fBm images analysed in this work, grouped by common phase distribution
and color-coded as in Figure~\ref{fbmmfs} (from dark to light as the power
spectrum exponent $\beta$ increases from 2 to 4 in steps of 0.4).}\label{allfbmdq}
\end{figure*}

To complete this comparative analysis, we also examine the behaviour of $D_q$ vs $q$ also for the
simulated column density maps and the fBm sets considered in this paper (Figures~\ref{dqmodel} 
and~\ref{allfbmdq}, respectively). 

In the case of simulations, arguments similar to those emerging from 
the MFS analysis can be repeated: the elongation of $D_q$ values, both at $q<0$ and $q>0$, 
in the ``compressive forcing'' case 
is larger than in the ``solenoidal forcing'' one (Figure~\ref{dqmodel}, top panels), connected to the 
fact that the MFS is found to be broader. The very low $D_q$ values at positive $q$ orders for the maps 
corresponding to the $t_2$ epochs for the ``quasi-hydrodynamical'' and ``high-magnetization'' scenarios 
(Figure~\ref{dqmodel}, bottom panels) are parallel to the broadening of the MFS left tail as a consequence 
of the increase of star forming activity in these simulations.

The fBm images show a strong divergence between the negative $q$ portion of the $D_q$ curve, which 
gets higher as $q$ decreases or $\beta$ increases, and the positive $q$ one, in which differences 
between images at different $\beta$ are much smaller (although still systematic). Again, this effect
corresponds to the strong broadening of the right tail of the MFS, with respect to the left one, 
observed at increasing $\beta$ (Figure~\ref{allfbmmfs}).   

\begin{figure}
\includegraphics[width=8.5cm]{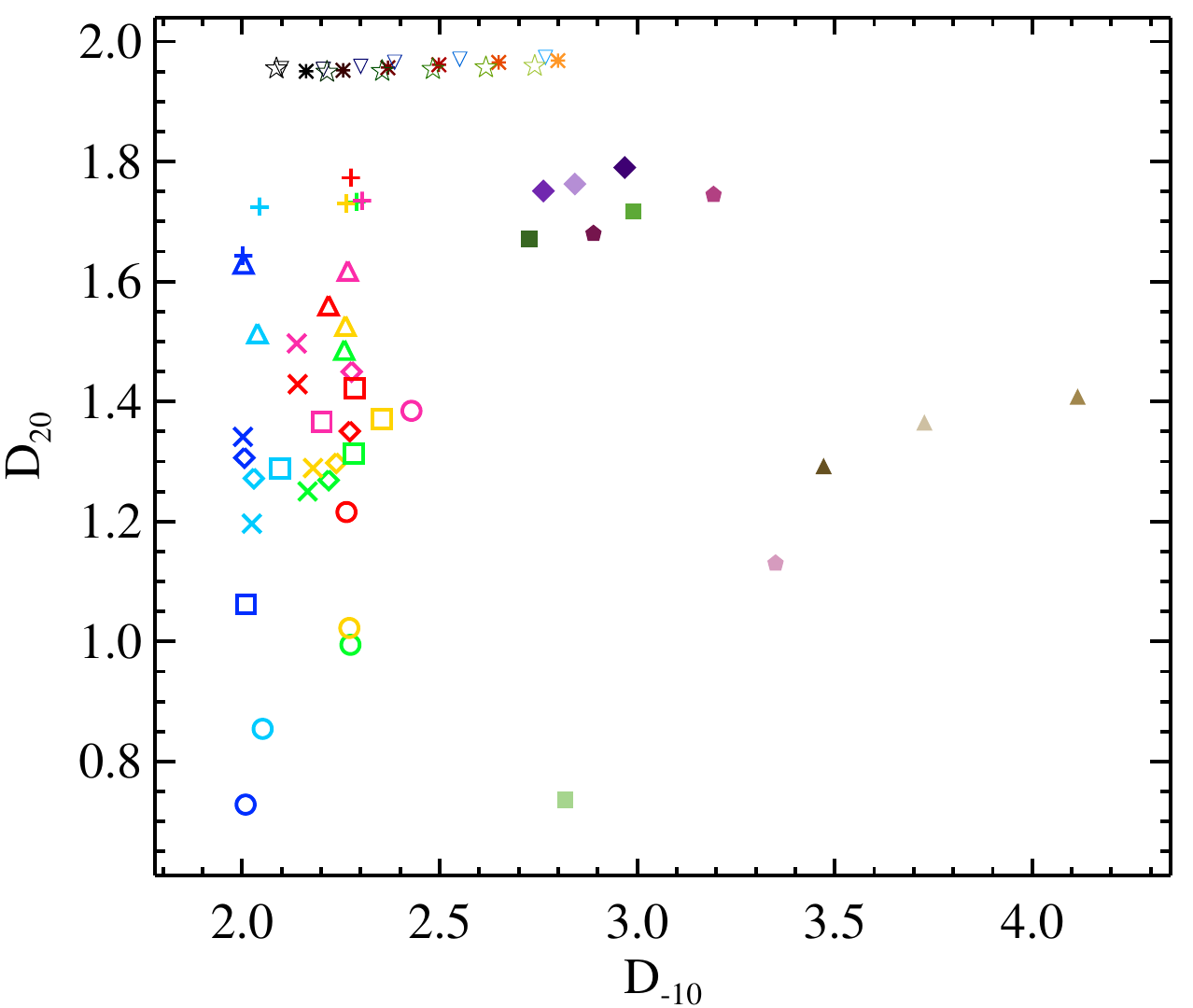} \caption{Fractal dimension of order $q=20$ versus 
the one of order $q=-10$ for all the maps analysed in this paper. Symbols and colors are the same
introduced in Figure~\ref{chappellplot}.}\label{dimextent}
\end{figure}

On the basis of the above considerations, a plot based on salient features of the $D_q$ curves should not 
add new information with respect to the analysis in Section~\ref{quantifying} using the main features of the 
MFS. Nevertheless, since the $D_q$ curves are not affected by features like the cusps appearing
in some MFS, there is no limitation, for example, in displaying the $D_{20}$ versus $D_{-10}$
for all the investigated maps, as a measure of the excursion between values found at the extreme
positive and negative probed orders $q$ (Figure~\ref{dimextent}). The three classes of images
populate different zones of the diagram:
\begin{itemize}
\item The Hi-GAL images occupy a vertical belt delimited by $2.0 \lesssim D_{-10} \lesssim 2.4$, in which,
from left to right, the PACS maps are encountered before the SPIRE and column density ones. At the top
of the group, the $\ell215$ and $\ell222$ maps are found, i.e. the most quiescent fields, while the 
$\ell217$ maps represent the bottom of the population.
\item The numerical simulations occupy a wide region of the plot at $D_{-10} \gtrsim 2.7$, in which
it is possible to distinguish two further sub-groups: ($i$) seven maps on the top ($D_{20}>1.6$) 
including the three of the ``solenoidal'' case, and the maps (on both projections) at $t_1$ of the 
``quasi-hydrodynamical'' and ``high-magnetization'' cases, and ($ii$) five maps at lower $D_{20}$ 
including the three of the ``compressive forcing'' case, and the maps at $t_2$ of the 
``quasi-hydrodynamical'' and ``high-magnetization'' cases.
The seven maps of the group $i$ correspond to statistically similar structure, shaped by 
pure turbulence. When gravity, absent in the evolution of the ``solenoidal'' scenario, becomes 
predominant at $t_2$ in the ``quasi-hydrodynamical'' and ``high-magnetization'' cases, $D_{20}$
decreases, more significantly in the first case, in absence of magnetic support of the cloud
(see Appendix~\ref{simevol} for a more systematic analysis of the effects of gravity and magnetic
field).
\item A narrow horizontal belt at $D_{20}= 1.95$ populated by fBm images, being the low-$\beta$
on the left and the high-$\beta$ on the right.
\end{itemize}

Figure~\ref{dimextent} summarizes some of the main results of our analysis: at least for our choice of 
observations and simulations, the multifractal behaviour of the former and of the latter
appear quite different, despite the wide range of physical conditions explored. Qualitatively speaking,
common behaviours can be recognised, such as the decrease of $D_{20}$ (or, correspondingly, the broadening 
of the left tail of the MFS) as the star forming content increases, but from the quantitative point of view
the typical values of generalised fractal dimensions remain remarkably different, so that a clear
segregation of image classes is found. The fBm images denote again a different behaviour with
respect to the other two classes, and, in the light of the multifractal analysis, seem inadequate to 
be representative of the statistical properties of ISM, contrarily to what the mono-fractal analysis
typically suggests.

\section{Conclusions}\label{conclusions}
The need of extending the multifractal description of the ISM, so far attempted only by \citet{cha01},
to more recent, high-quality, and statistically meaningful data motivated this work. We carried out 
the multifractal analysis of six Hi-GAL fields in a region of the third Galactic quadrant already
described by \citet{eli13}, and characterised by a conveniently low degree of 
velocity component overlap along the line of sight. These fields are interesting by themselves,
thanks to the variety of morphologies and physical conditions displayed by the ISM in this region. 

For comparison, the same analysis was extended to 
30 synthetic maps: 12 column density maps of simulated clouds extracted from the STARFORMAT data base
and obtained from four different numerical models, at two projections and two evolutionary epochs, 
and 18 fBm sets spanning a variety of power spectrum slopes and initial phase distributions.

A number of results emerged from this analysis.
The most important ones are summarised as follows:

\begin{enumerate}
\item All the investigated fields exhibit a multifractal rather than a simple monofractal structure. 
The variety of the obtained MFS shapes permits a further differentiation among them. 
\item Strong differences are found between the left and the right tail of the MFS. Most of the
analysed Hi-GAL maps exhibit a broader left tail of the MFS with respect to the right one, with the 
extreme case of the 70~$\mu$m maps, presenting a right tail almost collapsed to a single point 
against a pronounced left one, signature of the presence of several bright spots on a very low-signal 
background, as the maps at this wavelength typically appear in this portion of the Galaxy.
\item The roughly linear relation between the strength of the brightest singularities and 
dimensional diversity in the ISM maps found by \citet{cha01} is confirmed by our analysis
of Hi-GAL fields at various wavelengths. 
\item The peak position and the total width of the MFS appear to systematically drift, for
a given tile, with wavelength, denoting an increase of image complexity at increasing wavelength
due to the progressive appearance of a network of filaments and of cold structures
in general. Moreover, global behaviours are recognisable from tile to tile, from
star formation-rich to more quiescent ones (more and less ``complex'', respectively).
\item The MFS of the Hi-GAL fields is generally left-skewed, with the exception of the $\ell215$ 
tile. This is mostly associated to a lower $f$ initial value for the predominant tail with
respect to the opposite extreme of the MFS, so that the ``vertical'' versus ``horizontal balance'' 
of the MFS appear to follow a direct power law with exponent close to 1.
\item Comparing the generalised fractal dimensions of the Hi-GAL maps with the mono-fractal ones,
a clear correlation is seen at positive orders (nearly linear at $q=10$), suggesting that
the estimates of the fractal dimension are strongly influenced by the strongest brightness
peaks present in the maps. 
\item The cloud models partly exhibit different behaviours, depending on the underlying physics. The 
``compressive forcing'' maps show a different MFS compared to observations. The ``solenoidal 
forcing'', ``quasi-hydrodynamical'', and ``high-magnetization'' show relatively similar spectra at 
the earliest of the two considered epochs (with small differences among different projections), but 
with the last two presenting a systematic broadening of the left tail of the MFS as their 
star formation content increases
under gravity action, in agreement with the behaviour recognised for observational maps.
The presence of a magnetic support of the cloud, instead, is seen to contrast this effect
\item Despite qualitative similarities between the MFS derived for observed maps and numerical simulations,
using quantitative descriptors of the MFS shape (such as peak position, spectrum width and skewness, 
etc.) a segregation between the two image classes is found. This is also emphasized
by the analysis of the generalised fractal dimension, suggesting that, at least for the
analysed set of maps, simulations do not completely fulfill the constraints
on the structure suggested by the multifractal analysis of observational maps.
\item The multifractal analysis of fBm images reveals that they show a remarkably different behaviour 
compared with our observed maps, suggesting that, despite analogies between
power spectra, the Fourier phase distributions for these two classes can be quite different and
responsible of the observed discrepancies. All the used indicators advise against the 
use of the fBm images as a surrogate for the observations.
\end{enumerate}

Moreover, this study has produced some interesting results about fBm images by themselves, 
of interdisciplinary relevance independently from the comparison with observations: 
\begin{enumerate}
\setcounter{enumi}{9}
\item The increase of the power spectrum exponent $\beta$ generally produces a rise of the last point of 
the left part of the MFS and a strong systematic enlargement of the right tail of the MFS. The peak position
increases as well at increasing $\beta$.
\item On the other hand, the MFSs of two fBm generated with the same $\beta$ and a different random phase 
distribution, are not identical although they have the same fractal dimension. This represents a 
further evidence of the general importance of analysing the Fourier phases together with the simple
power spectrum.
\end{enumerate}

Possible extension of this work will consist in enlarging the sample of analysed \textit{Herschel} fields,
provided that they are associated to a predominant gas velocity component, as well as the number
of the considered models, searching for possible points of contact of the two classes that could have been missed
in the present analysis.

\section*{Acknowledgements}
NS acknowledges support by the French ANR and the German DFG through the project ``GENESIS''
(ANR-16-CE92-0035-01/DFG1591/2-1). 
ES acknowledges support from the EU-funded VIALACTEA programme (FP7-SPACE-607380607380).
Herschel is an ESA space observatory with
science instruments provided by European-led Principal Investigator
consortia and with important participation from NASA.

%The Acknowledgements section is not numbered. Here you can thank helpful
%colleagues, acknowledge funding agencies, telescopes and facilities used etc.

%%%%%%%%%%%%%%%%%%%%%%%%%%%%%%%%%%%%%%%%%%%%%%%%%%

 

\bibliographystyle{mnras}
\bibliography{fractbib}

%%%%%%%%%%%%%%%%% APPENDICES %%%%%%%%%%%%%%%%%%%%%

\appendix

\section{fBm images used in this analysis}\label{fbmappend}
\begin{figure*}
\includegraphics[width=15cm]{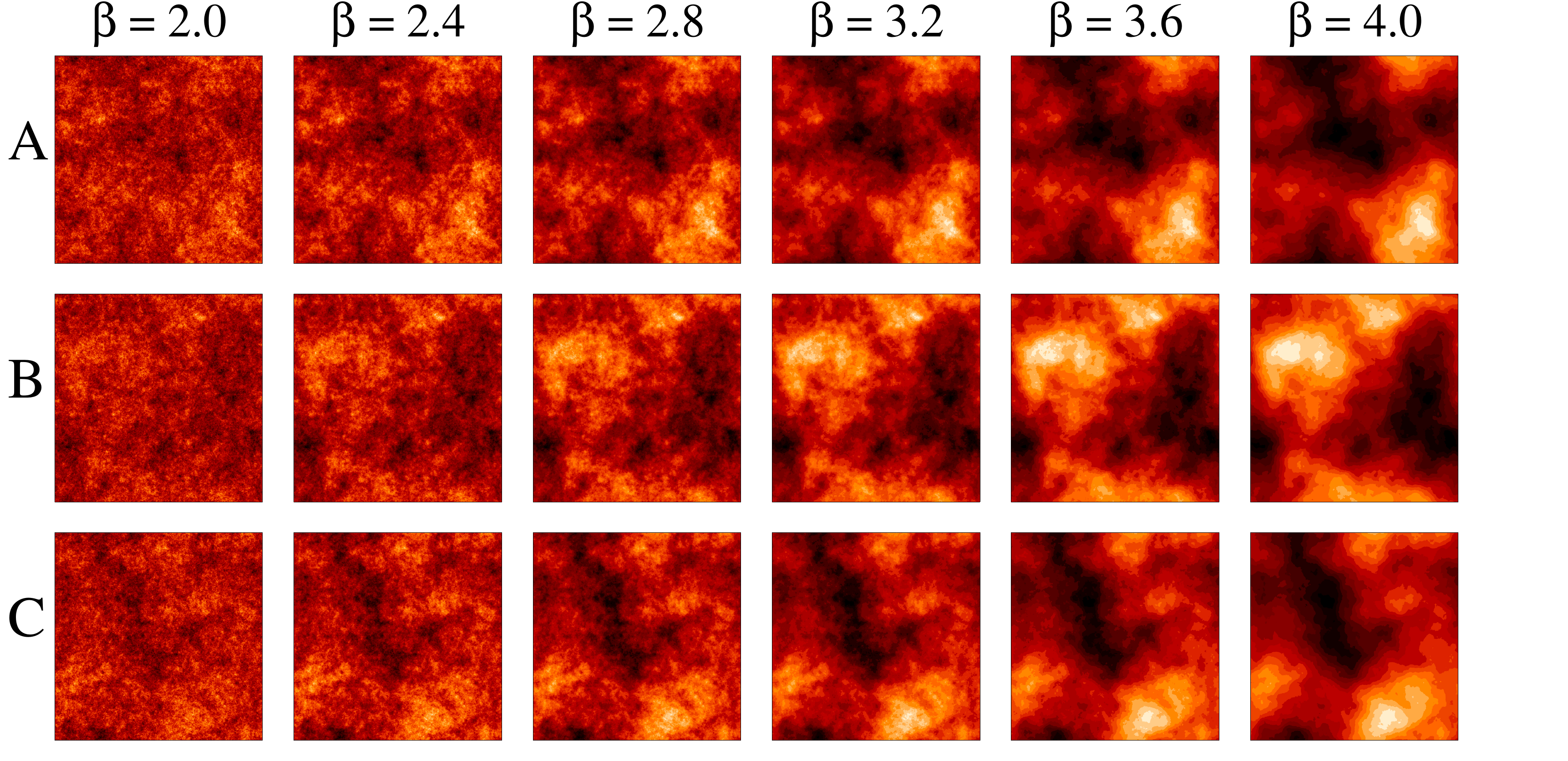} \caption{fBm $1000\times 1000$ pixel images generated 
as described in Section~\ref{fbm}. Images are organized in rows according to the three 
different (A, B, and~C) phase distributions adopted, and in columns according to the exponent 
$\beta$ of their power-law power spectrum, which is reported on the top (ranging from 2 to 4 in steps 
of 0.4).}
\label{fbmfig}
\end{figure*}

The 18 fBm images generated, by varying the power-law power spectrum exponent and the random phase distribution 
of their Fourier transform, as reference sets for the multifractal analysis carried out in this paper
are shown in Figure~\ref{fbmfig}.

%\section{    }\label{tenvsfive}

\section{Evolution in time of multifractal parameters of simulations}\label{simevol}
In this appendix we expand the analysis of the MFS of numerical simulations with respect
to the evolution of models with time, to assess if the trends recognised in Section~\ref{results},
and in particular the broadening of the left tail of the MFS in simulations containing gravity, can 
be seen as genuine, or as simple fluctuations before the convergence is reached 
\citep[cf., e.g.,][]{syt00,fed10}.

For this purpose, it is necessary to follow the MFS evolution over a number of time steps larger than 
two. Browsing the STARFORMAT data base, one finds that the ``solenoidal forcing'' and ``compressive forcing''
scenarios are sampled with nine time snapshots (of which we analysed the second and the eighth one in the main
article), while for the ``quasi-hydrodynamical'', and ``high-magnetization'' only two time snapshots 
are available in all, coinciding with those analysed so far. To overcome this issue for these two simulations, 
on the one hand we expanded the number of analysed maps including all possible scenarios and projection
directions, and on the other hand we used further simulations, as described below.

Among the geometrical features of the MFS, we here focus in particular on the width of the left tail, 
which can be expressed through $\Delta \alpha_L$ (see Section~\ref{mfswidths}), and on the one of the 
right tail ($\Delta \alpha_R$), used for comparison. However, as extensively explained in Section~\ref{results},
the values of $\alpha_{10}$ and $\alpha_{-10}$ used, respectively, to derive those two quantities can
be affected by the issue of cusp-like features in the MFS. Instead, as discussed in Section~\ref{dqanalysis},
the values of the generalised fractal dimension $D_q$ do not suffer from this problem and, at the same time,
offer a description which is completely equivalent to the one based on $\Delta \alpha_L$ and $\Delta \alpha_R$.
Indeed, a broadening of the left tail corresponds to a decrease of $D_q$ at positive $q$ orders,
while for the right tail it corresponds to an increase of $D_q$ for negative $q$ (see, e.g.,  
Figure~\ref{dqmodel}). For this reason, we here opt for a description based on the evolution with time
of the $D_{20}$ and $D_{-10}$ parameters (already used also in Figure~\ref{dimextent}), sensitive to
large and small fluctuations in the image, respectively. 

\begin{figure}
\includegraphics[width=8.5cm]{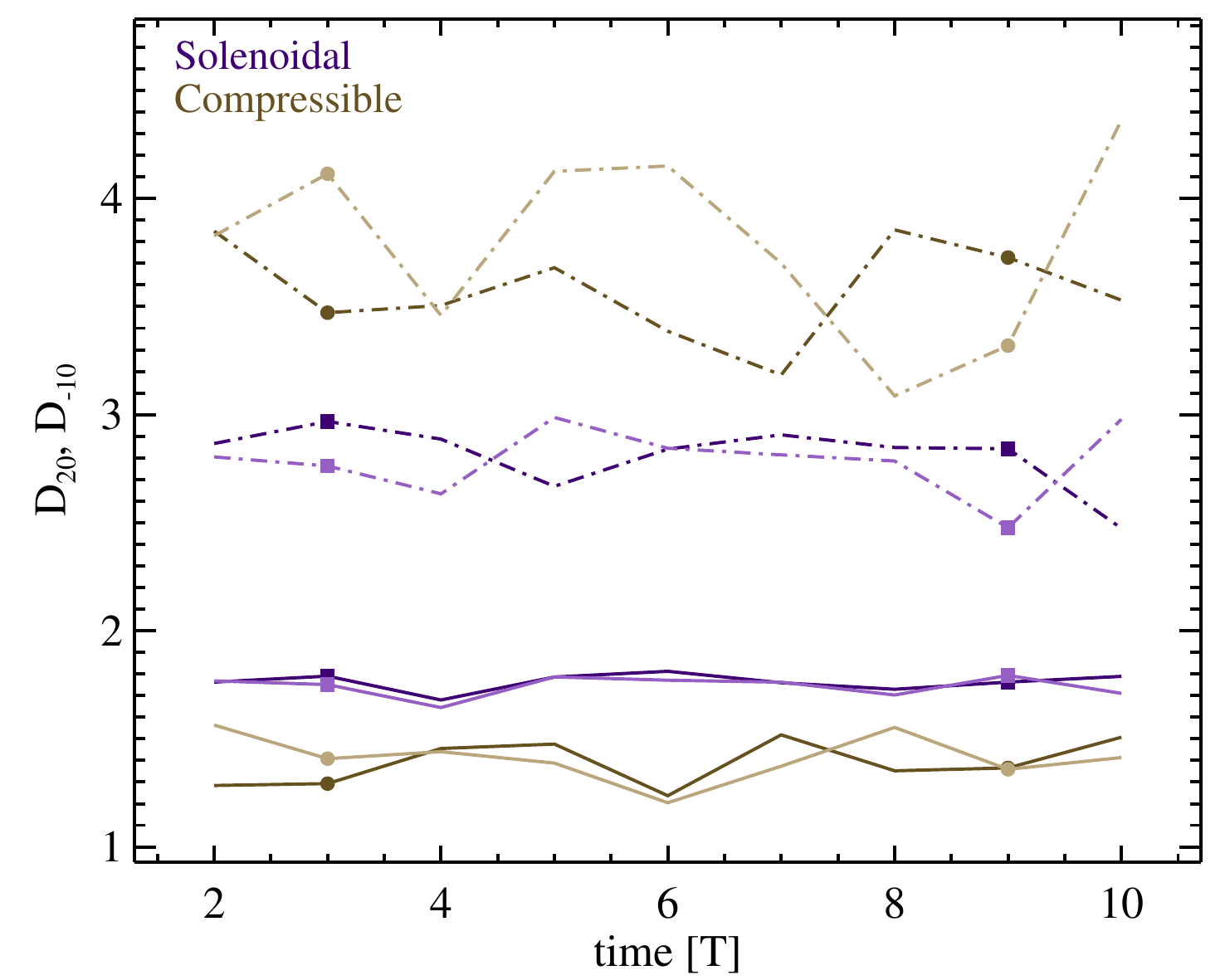} \caption{Evolution in time of the fractal dimensions $D_{20}$ (solid lines, lower part of the plot) and $D_{-10}$ (dotted-dashed lines, upper part) for the 
``solenoidal forcing'' (purple) and ``compressive forcing'' (brown) numerical simulations of
column density, respectively. The two  projections $yz$ and $xy$
(see Section~\ref{simulations}) are taken into account (darker and lighter color, respectively). The time 
is in units of $\mathcal{T}$ (see Section~\ref{simulations}), and the values of $D_{20}$ and $D_{-10}$ for
$\mathcal{T}=3$ and $9$ are marked with symbols (assigned as in Figure~\ref{chappellplot}), as they 
coincide with those already calculated and discussed in Section~\ref{dqanalysis}.}
\label{federrath_time}
\end{figure}

These two dimensions have been computed for all time steps available for the ``solenoidal forcing'' 
and ``compressive forcing'' scenarios, and are shown in Figure~\ref{federrath_time}.  No monotonic
trend is found for both the scenarios and for both considered projections. In this respect, as 
already highlighted 
in Section~\ref{results}, the $D_{-10}$ values appear more spread than the $D_{20}$ ones.

\begin{figure}
\includegraphics[width=8.5cm]{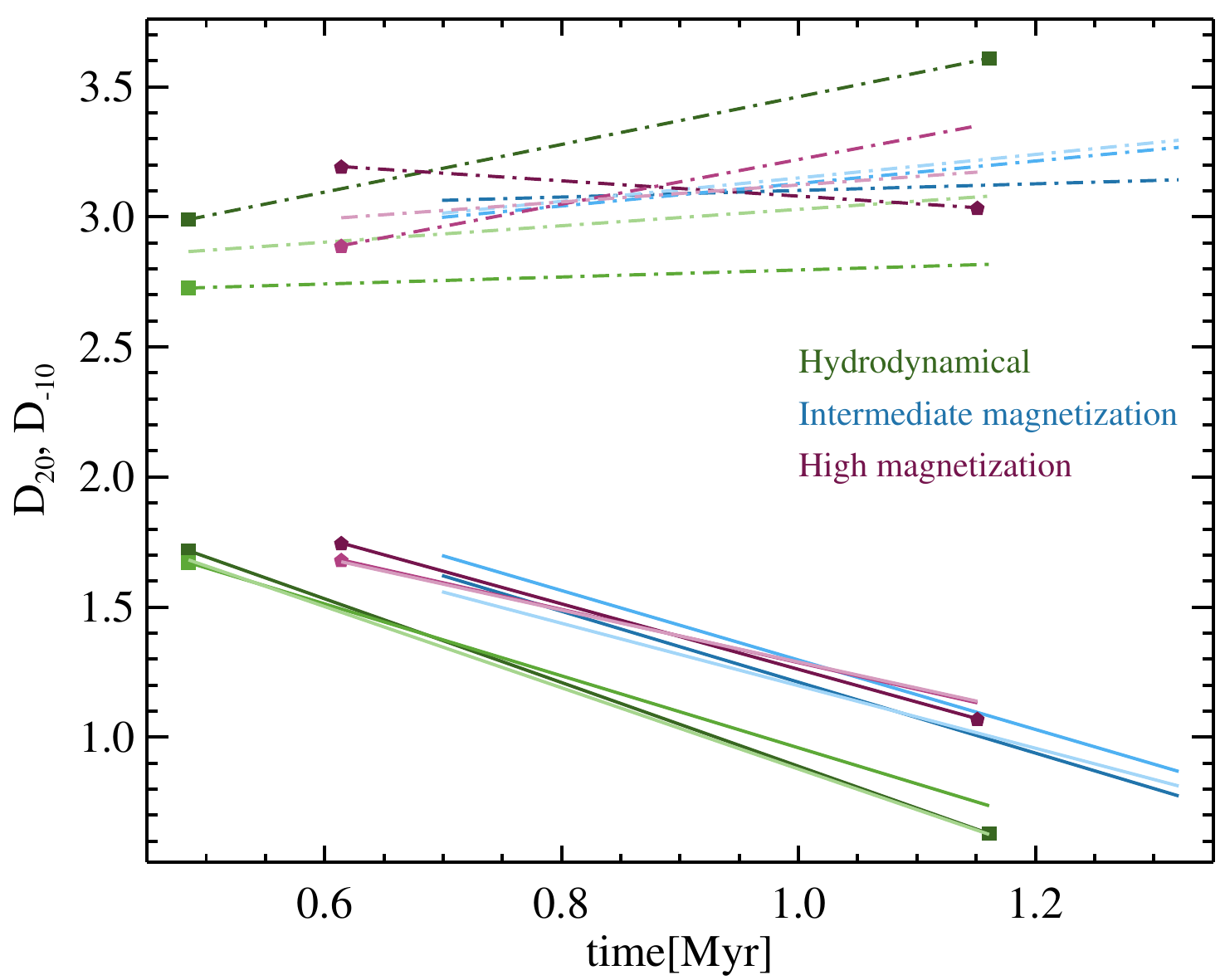} \caption{Evolution in time of the fractal dimensions 
$D_{20}$ (solid lines, lower part of the plot) and $D_{-10}$ (dotted-dashed lines, upper part) for the 
``quasi-hydrodynamical'' (green), ``intermediate magnetization'' (blue), and ``high magnetization'' 
(pink) numerical simulations of column density, respectively. All three projections are taken into account 
(dark, intermediate and and lighter color for $yz$, $xy$, and $xz$, respectively). Only two time 
steps are available, with the two times depending on the scenario (cf. Section~\ref{simulations}). As 
in Figure~\ref{federrath_time}, the values of $D_{20}$ and $D_{-10}$ already calculated and shown in
Section~\ref{dqanalysis} for particular combinations of scenarios and projection directions are marked 
with symbols (assigned as in Figure~\ref{chappellplot}).}
\label{starformat_time}
\end{figure}

As said above, for the other two scenarios, only two time steps $t_1$ and $t_2$ are available. 
Thus to increase statistics, we added to our analysis the third projection direction available, 
namely $xz$, and a further ``intermediate magnetization'' scenario available on STARFORMAT, 
characterized by equal values for thermal and magnetic pressure (images can be found in the STARFORMAT 
web site). Figure~\ref{starformat_time} shows the evolution from $t_1$ to $t_2$ of the $D_{20}$ and $D_{-10}$ 
dimensions for the three scenarios, considering for each of them the three possible projections.
Although based only on two time steps, this analysis shows a systematic decrease of $D_{20}$ with time
(equivalent to the broadening of the MFS left tail) for all scenarios and projection directions,
explainable with the enhancement of singularities, both in number and intensity, in the maps due 
to the action of gravity. This is further confirmed by the fact, already suggested in Section~\ref{results}, 
that this observed 
trend for $D_{20}$ is stronger as the magnetic support of the cloud implied by the model decreases. 
Here, considering an additional intermediate scenario, we can further probe this hypothesis: for the 
``quasi-hydrodynamical'' case, the slopes of the segment are -1.61, -1.38, and -1.56 for the $xy$, $yz$, 
and $xz$ projections, respectively; for the ``intermediate magnetization'' case, -1.36, -1.33, and -1.20;
for the ``high magnetization'' case, -1.25, -1.02, and -1.00. Therefore, we can summarize that, 
within a certain degree of 
fluctuation due to projection effects, the MFS at large positive $q$ shows a broadening with time due
to gravity, but mitigated by the possible presence of the magnetic field. Finally, we notice an almost 
systematic (in eight out of nine total cases) increase of $D_{-10}$ with time (equivalent to the 
broadening of the right tail of the MFS), already highlighted in Section~\ref{results}. This is probably
due to the appearance of small cavities in the maps as a result of the gravity action (see
Figure~\ref{modelimage}, panels for time $t_2$ of last two scenarios, especially the first of the two).

\begin{figure}
\includegraphics[width=8.0cm]{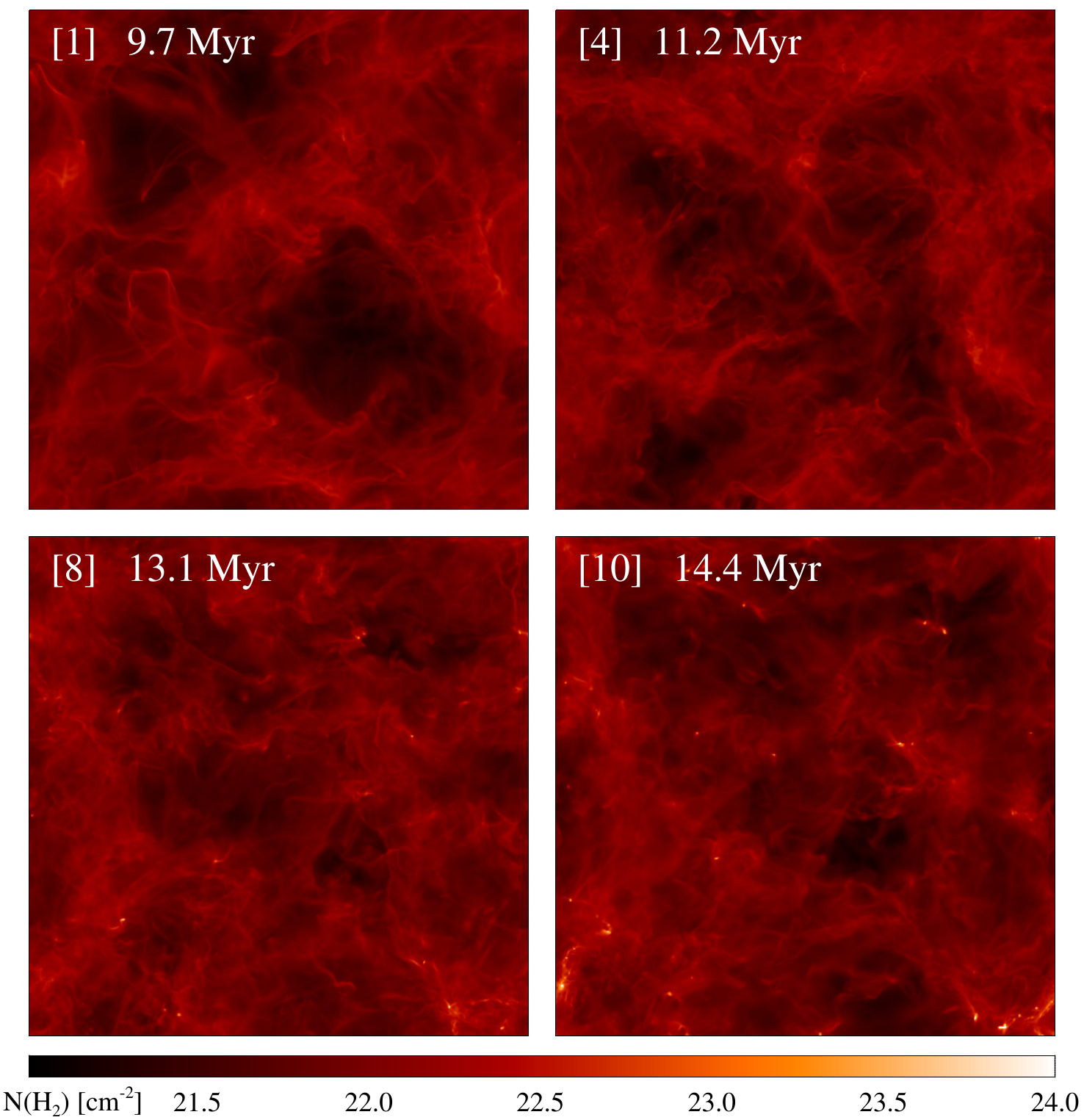} \caption{Four (out of ten) examples of 
column density snapshots obtained for this paper similarly to \citet{hen18}. The number of the 
snapshot and the corresponding time are overwritten in white in each panel. In particular,
the initial (top left) and final (bottom right) snapshots were chosen and displayed, and the
first after the gravity is switched on (top right). Maps are in decimal logarithm of column 
density (in cm$^{-2}$). The bottom color bar applies to all maps, and to improve the rendering 
of relatively faint structures, it saturates at a value smaller than the maximum reached in the 
bottom right panel, namely $\sim 25.2$.}
\label{patrick_images}
\end{figure}

\begin{figure}
\includegraphics[width=8.5cm]{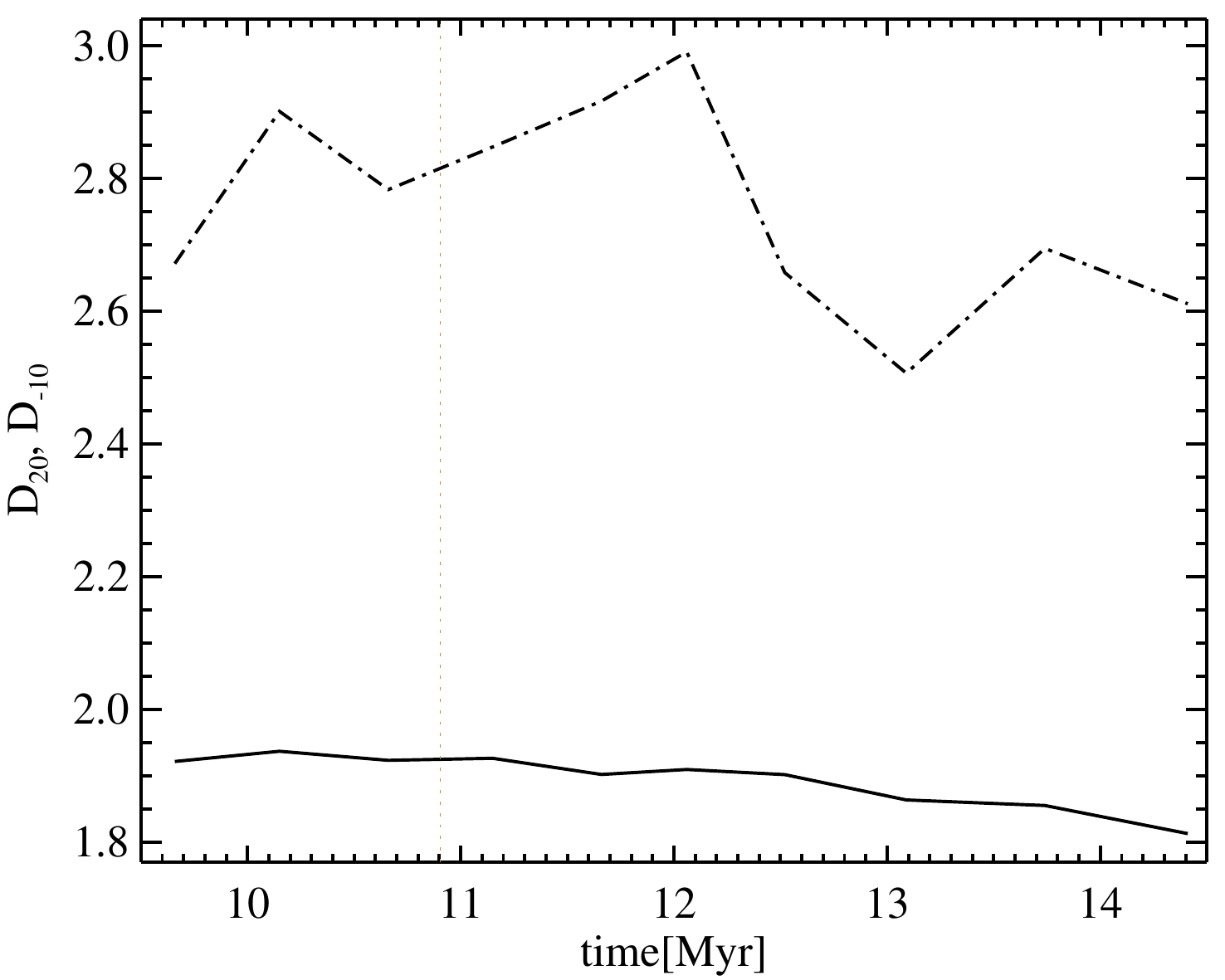} \caption{Evolution of the fractal dimensions 
$D_{20}$ (solid line, lower part of the plot) and $D_{-10}$ (dotted-dashed line, upper part) for the 
simulations carried out for this analysis, based on the approach of \citet{hen18} and partially
shown in Figure~\ref{patrick_images}. The dotted vertical 
grey line indicates the middle point between the last time step with gravity switched off and the first
one with gravity switched on.}
\label{patrick_time}
\end{figure}

To ascertain if the observed evolution with time of the MFS can be considered genuine, one needs an analysis 
based on several time steps. For this reason we analysed two further sets of simulations involving 
gravity. The first one, obtained on purpose for this work, is based on the same approach used in
\citet{hen18}. Ten $1024^2$ pixel column density snapshots were obtained. Only driven turbulence is 
present at the beginning (initial density of 200~cm$^{-3}$, Mach number $\sim 10$), whereas gravity 
is switched on between the third and the fourth step. The MFS and the set of generalised fractal 
dimensions were computed. For the sake of brevity, here we show directly the plot of $D_{20}$ and 
$D_{-10}$ vs the time (Figure~\ref{patrick_time}): while the latter do not show any particular trend 
with time, the former decreases after the gravity is switched on and compact structures start to 
appear in the maps.

\section{$\Delta$-variance analysis of Hi-GAL fields}\label{dvarapp}
In this work, focused on the multifractal analysis of the ISM, we also intend to compare multifractal
parameters of the six investigated fields with the ``classic'' (mono-)fractal dimension (see 
Section~\ref{dqanalysis}), derived through the $\Delta$-variance technique \citep{stu98}, as in 
\citetalias{eli14}. However it has been necessary to repeat the $\Delta$-variance calculations here, 
both because of the addition of two new fields with respect to \citetalias{eli14}, namely $\ell215$
and $\ell226$, and of the fact that, also for the other four tiles in common with \citetalias{eli14},
the frames analysed in this work are slightly different in size and position, as explained in
Section~\ref{higal}.

More details, both theoretical and operational, about the $\Delta$-variance analysis are given 
in \citetalias{eli14}. Here we simply summarise that, in the case of a power-law power spectrum set 
(such as a fBm image), this technique yields a robust estimate of the exponent $\beta$ of such
spectrum without resorting to Fourier analysis. The average of the convolution of the image with a 
Mexican hat function with a characteristic scale length $L$ gives the value $\sigma^2_{\Delta}$ of 
the $\Delta$-variance for this length. \citet{stu98} demonstrated that, for a fBm image,
\begin{equation}
\sigma^2_{\Delta}(L)\propto L^{2-\beta}\,.
\end{equation}
A natural fractal image, instead, can exhibit such a 
behaviour on a limited range of scales (or on multiple ranges, with different slopes) on which 
it appears self-similar\footnote{As in \citetalias{eli14}, here for each tile we identify an
appropriate range on which all wavelengths show an approximately power-law behaviour.}.

\begin{figure*}
\includegraphics[width=15cm]{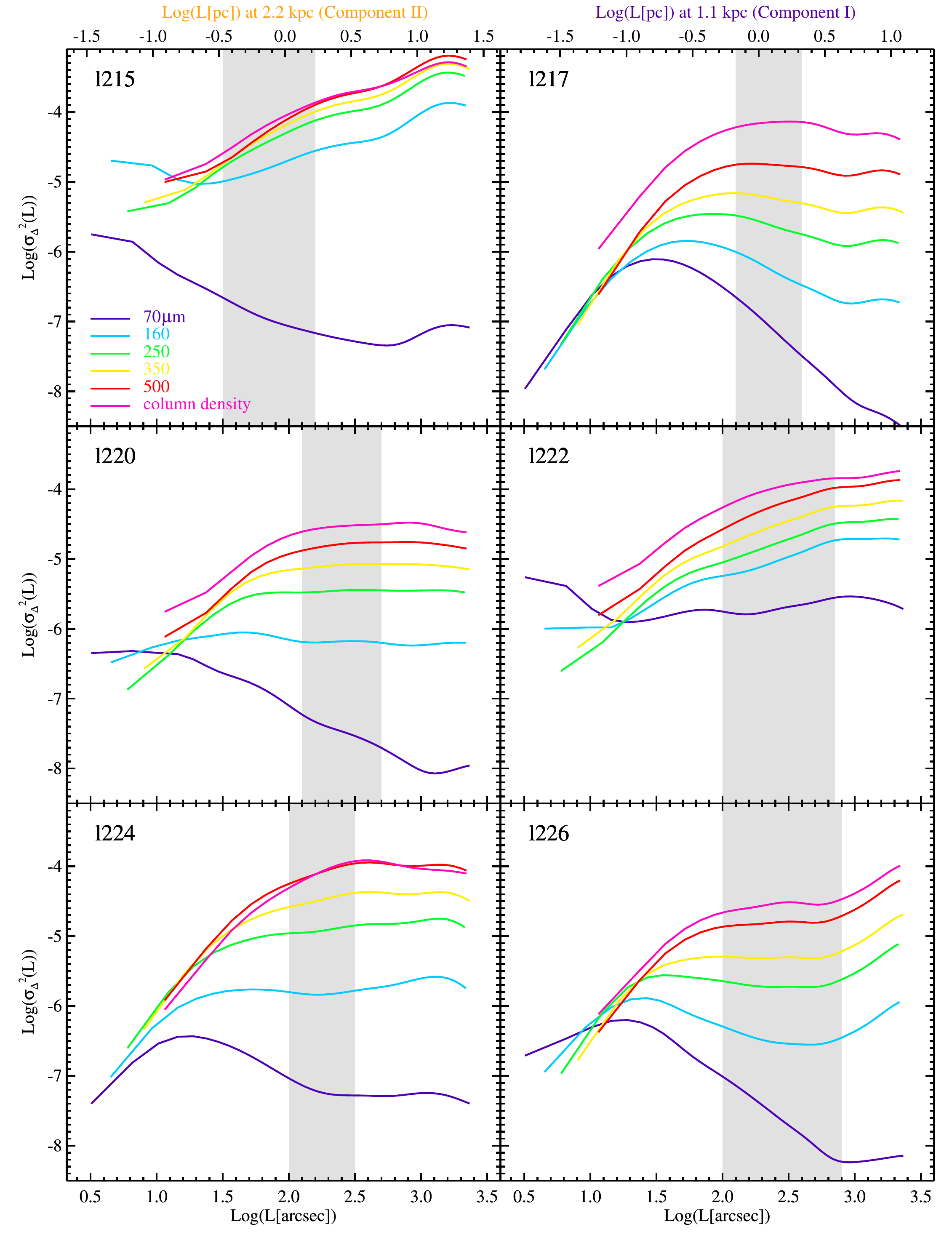} \caption{$\Delta$-variance curves of the maps 
shown in Figure~\ref{htiles} using the same tile naming and band-color encoding (this figure is 
similar to Figure~4 of \citetalias{eli14}). Tiles associated
to velocity component~II are in the left panels, those associated with component I are in the 
right panels (for reference, the spatial scales corresponding to angular scales at the distances
of the two components are reported on the top axis of the two upper panels, respectively). We 
use the same axis ranges in all panels to allow a direct comparison of the slopes. The inertial 
range is highlighted as a grey area in each panel. The corresponding linear
slopes are transformed in power spectrum slopes (see text) are reported in Table~\ref{dvartab}.}
\label{dvarfig}
\end{figure*}

\begin{table*}
\centering
\caption{Power spectrum exponent and fractal dimension of the investigated Hi-GAL maps \label{dvartab}}
\begin{tabular}{lccccccccccccc}
\hline
 {Field} & Distance  & Fit range & \multicolumn{5}{c}{$\beta$} & & \multicolumn{5}{c}{$D$} \\
\cline{4-8} 
\cline{10-14} \\
 & [kpc] & [pc] & 160 $\mu$m & 250 $\mu$m & 350 $\mu$m & 500 $\mu$m & Col. dens. & & 160 $\mu$m & 250 $\mu$m & 350 $\mu$m & 500 $\mu$m & Col. dens. \\
\hline
$\ell$215 & 2.2 & 0.3--1.7 & 2.65 & 2.94 & 3.05 & 3.21 & 3.02 &  & 2.67 & 2.53 & 2.47 & 2.40 & 2.49\\
$\ell$217 & 2.2 & 1.3--4.2 & 0.99 & 1.44 & 1.67 & 1.92 & 2.17 &  & 3.50 & 3.28 & 3.17 & 3.04 & 2.92\\
$\ell$220 & 2.2 & 1.3--5.3 & 2.03 & 2.07 & 2.11 & 2.20 & 2.16 &  & 2.99 & 2.97 & 2.95 & 2.90 & 2.92\\
$\ell$222 & 1.1 & 0.5--3.8 & 2.65 & 2.67 & 2.68 & 2.69 & 2.47 &  & 2.67 & 2.66 & 2.66 & 2.66 & 2.76\\
$\ell$224 & 1.1 & 0.5--1.7 & 2.05 & 2.25 & 2.43 & 2.62 & 2.82 &  & 2.97 & 2.87 & 2.78 & 2.69 & 2.59\\
$\ell$226 & 1.1 & 0.5--4.2 & 1.81 & 1.99 & 2.02 & 2.10 & 2.14 &  & 3.09 & 3.00 & 2.99 & 2.95 & 2.93\\
\hline
\end{tabular}
\end{table*}

Similarly to \citetalias{eli14}, in Figure~\ref{dvarfig} the $\Delta$-variance curves of the six 
Hi-GAL fields (in five bands, plus the corresponding column density maps) as a function of
$L$ are shown, with the indication of the spatial scale range identified for the determination of 
$\beta$ through linear best fit of the logarithms. The fit range for the four central tiles has been 
kept identical to \citetalias{eli14}, whereas for the tiles introduced in this paper, namely $\ell215$ 
and $\ell226$, it has been estimated as described in \citetalias{eli14}. In Table~\ref{dvartab} the 
obtained values of $\beta$, together with the corresponding fractal dimension $D_\Delta$ (according 
to Equation~\ref{dfbm}) are reported.

%%%%%%%%%%%%%%%%%%%%%%%%%%%%%%%%%%%%%%%%%%%%%%%%%%

% Don't change these lines
\bsp	% typesetting comment
\label{lastpage}
\end{document}